\documentclass[fleqn,10pt]{wlscirep}
\usepackage[utf8]{inputenc}
\usepackage[T1]{fontenc}

\usepackage{import}
\usepackage{multirow}

\usepackage{chemformula}
\usepackage[separate-uncertainty = true,multi-part-units=single]{siunitx}
\DeclareSIUnit[]{\bohrmagneton}{\text{mb}}
\DeclareSIUnit[]{\amu}{\text{u}}

\usepackage{physics}

\usepackage{layouts}

\usepackage{cleveref}

\usepackage{comment}

\usepackage{interval}
\intervalconfig{
                soft open fences ,
                separator symbol =; ,
                }

\usepackage{subcaption}

\usepackage{ulem}

\usepackage{glossaries}
\glsdisablehyper
\newacronym{srf}{SRF}{superconducting radio frequency}
\newacronym{linac}{linac}{linear accelerator}
\newacronym{rf}{RF}{radio frequency}
\newacronym{cw}{CW}{continuous wave}
\newacronym{Rs}{\ensuremath{R_s}}{RF surface resistance}
\newacronym{Q0}{\ensuremath{Q_0}}{quality factor}
\newacronym{Ea}{\ensuremath{E_\mathrm{acc}}}{accelerating gradient}

\newacronym{b-c1}{\ensuremath{B_{c1}}}{lower critical field}
\newcommand{\BcOne}{\ensuremath{B_{c1}}}

\newacronym{b-c2}{\ensuremath{B_{c2}}}{upper critical field}
\newcommand{\BcTwo}{\ensuremath{B_{c2}}}

\newacronym{b-c}{\ensuremath{B_\text{c}}}{thermodynamic critical field}

\newacronym{b-sh}{\ensuremath{B_\mathrm{sh}}}{superheating field}

\newacronym{b-vp}{\ensuremath{B_{vp}}}{field of first vortex penetration, $\approx$180 mT at 0K}

\newcommand{\BApp}{B_\text{a}}
\newacronym{b-app}{\ensuremath{B_\text{a}}}{applied field}

\newacronym{b-gap}{\ensuremath{B_{g}}}{gapless superconducting field}

\newacronym{tc}{\ensuremath{T_c}}{superconducting critical temperature}
\newacronym{tc2}{\ensuremath{T_{c2}}}{superconducting critical temperature}

\newacronym{jv-c}{\ensuremath{\mathcal{J}_\text{c}}}{critical vortex current density}

\newacronym{RRR}{RRR}{residual resistivity ratio}
\newacronym{bcp}{BCP}{buffered chemical polishing}
\newacronym{ep}{EP}{electrolytic polishing}

\newacronym{mpd}{\ensuremath{\lambda(\lambda_L,\xi_0,l)}}{effective magnetic penetration depth}
\newcommand{\mpd}{\ensuremath{\lambda}}
\newacronym{mpd-zero}{\ensuremath{\lambda(\lambda_L,\xi_0,l)}}{effective magnetic penetration depth at $T=0$ K}

\newcommand{\mpdZeroZero}{\ensuremath{\mpd(0,0,\mfp)}}

\newacronym{bcs}{BCS}{Bardeen-Cooper-Schrieffer}
\newacronym{omegaGap}{\ensuremath{\Omega_G}}{energy gap for quasiparticle excitation}

\newacronym{delta-order-param}{\ensuremath{\Delta}}{superconducting order parameter or pair potential}

\newacronym{dos}{\ensuremath{N(\epsilon)}}{quasiparticle density of states}

\newacronym{xiZero}{\ensuremath{\xi_0}}{BCS coherence-length}
\newcommand{\xiZero}{\ensuremath{\xi_0}}

\newacronym{xiP}{\ensuremath{\xi_P}}{Pippard coherence-length}

\newacronym{lpd}{\ensuremath{\lambda_L}}{London penetration depth}
\newcommand{\lpd}{\ensuremath{\lambda_L}}

\newacronym{lpd-zero}{\ensuremath{\lambda_L}}{London penetration depth at $T=0$K}
\newcommand{\lpdZero}{\ensuremath{\lambda_L(T=0)}}

\newacronym{mfp}{\ensuremath{l}}{electron mean-free path}
\newcommand{\mfp}{\ensuremath{l}}

\newacronym{kappaGL}{\ensuremath{\kappa}}{Ginzburg-Landau parameter}
\newcommand{\kappaGL}{\ensuremath{\kappa}}

\newacronym{nlme}{NLME}{Nonlinear Meissner Effect}
\newacronym{lme}{LME}{Linear Meissner Effect}

\newacronym{zfc}{ZFC}{zero field cooled}

\newacronym{ims}{IMS}{Intermediate Mixed State}

\newacronym{sis}{SIS}{Superconductor-Insulator-Superconductor}

\newacronym{musr}{\ensuremath{\mu}SR}{muon spin rotation}
\newacronym{nmr}{NMR}{Nuclear Magnetic Resonance}
\newacronym{nqr}{NQR}{Nuclear Quadrupole Resonance}
\newacronym[sort={b-NMR}]{bnmr}{\ensuremath{\beta}-NMR}{\ensuremath{\beta}-detected nuclear magnetic resonance}
\newacronym[sort={muSR}]{le-musr-facility}{LE-\ensuremath{\mu}SR}{low-energy muon spin rotation facility at Paul Scherrer Institute}
\newacronym{xfel}{XFEL}{European X-ray Free Laser at DESY, Hamburg, Germany}
\newacronym{lcls2}{LCLS-II}{Linac Coherent Light Source II at SLAC, USA}

\newacronym{isac}{ISAC}{Isotope Separator and Accelerator}
\newacronym{isol}{ISOL}{isotope separation on-line}

\newacronym{slr}{SLR}{spin-lattice relaxation}

\newacronym{hv}{HV}{high voltage}
\newacronym{uhv}{UHV}{ultra-high vacuum}
\newacronym{ccd}{CCD}{charge-coupled device}

\newacronym{fem}{FEM}{finite element method}

\newcommand{\bNMR}{\ensuremath{\beta}\text{-NMR}\,}
\newcommand{\bNQR}{\ensuremath{\beta}\text{-NQR}\,}
\newcommand{\bSRF}{\ensuremath{\beta}\text{-SRF}\,}
\newacronym{le-musr}{LE-\ensuremath{\mu}SR}{low-energy muon spin rotation}

\newcommand{\LiEightPlus}{\ch{^{8}Li^{+}}\,}
\newcommand{\LiEight}{\ch{^{8}Li}\,}
\newcommand{\NbNinetyThree}{\ch{^{93}Nb}\,}
\newcommand{\MuPlus}{\ensuremath{\mu^+}\,}
\newcommand{\gLi}{\gamma_\text{\ch{^8 Li}}\,}
\newcommand{\gNb}{\gamma_\text{\ch{^{93} Nb}}\,}

\newcommand{\RelaxRate}{1/T_1\,}
\newcommand{\TOneNb}{\ensuremath{T_{1,\ch{Nb}}}}
\newcommand{\TOneBG}{\ensuremath{T_{1,\text{BG}}}}

\newacronym{rib}{RIB}{radioactive ion beam}

\newcommand{\NbBase}{“baseline”\,}
\newcommand{\NbMidT}{“mid-T bake”\,}
\newcommand{\NbMidTAlias}{“\ch{O}-doping”\,}
\newcommand{\NbNDope}{“\ch{N} doping”\,}

\newcommand{\NbLowT}{\SI{120}{\degreeCelsius} bake\,}
\newcommand{\NbTwoStep}{“two-step bake”\,}

\newacronym{hfqs}{HFQS}{“High Field Q Slope”}

\newacronym{tof-sims}{TOF-SIMS}{time-of-flight SIMS}
\newacronym{sims}{SIMS}{secondary-ion mass-spectrometry}
\newacronym{edx}{EDX}{energy dispersive X-ray spectroscopy}
\newacronym{sem}{SEM}{Scanning Electron Microscopy}
\newacronym{xps}{XPS}{X-ray photoelectron spectroscopy}

\newcommand{\aveBeam}[1]{\ensuremath{\left\langle #1 \right\rangle}}

\newacronym{chi-eff}{\ensuremath{\textit{e}(\BApp)}}{normalized enhancement factor}

\newcommand{\normEnhanceFact}{\ensuremath{\mathcal{N}}}

\newacronym{N-eff}{\ensuremath{\aveBeam{N_\text{eff,M}}}}{effective demagnetization factor averaged over beamspot area}

\newacronym{EnhFact-BeamAve}{\ensuremath{\mathcal{E}(\BApp)}}{enhancement factor}
\newcommand{\EnhanceFactBeamAve}{\ensuremath{\mathcal{E}(\BApp)}}
\newcommand{\EnhanceFactBeamAveNoBApp}{\ensuremath{\mathcal{E}}}

\newacronym{N-BeamAve}{\ensuremath{\aveBeam{N_\text{eff,M}}}}{field dependent effective demagnetization factor}

\newacronym{EnhFactMeissner}{\ensuremath{\aveBeam{\mathcal{E}}|_\text{Meissner}}}{average enhancement factor in the Meissner phase}

\newacronym{N-Meissner}{\ensuremath{\aveBeam{N_\text{eff,Meissner}}}}{effective demagnetization factor in the Meissner phase}

\newacronym{spd}{\ensuremath{\aveBeam{\Lambda}}}{experimental screening length}

\newacronym{b-surf}{\ensuremath{B_0}}{surface magnetic field}
\newcommand{\BSurf}{\ensuremath{B_0}}

\newacronym{b-int}{\ensuremath{B_{\rm in}}}{internal field}
\newcommand{\BInt}{\ensuremath{B_{\rm in}}}

\newacronym{b-entry}{\ensuremath{B_\text{entry}}}{applied field where flux is driven into the sample}
\newcommand{\BEntry}{\ensuremath{B_\text{entry}}}

\newcommand{\BMeissner}{\ensuremath{B_{\rm M}}}
\newcommand{\BVortex}{\ensuremath{B_{\rm v}}}

\newacronym{Impl-dist}{\ensuremath{\rho_E (x)}}{implantation stopping profile}
\newcommand{\ImplDist}{\ensuremath{\rho_E (x)}}

\newcommand{\Edward}{E.T.\,}
\newcommand{\Asad}{M.A.\,}
\newcommand{\Tobi}{T.J.\,}
\newcommand{\Philipp}{P.K.\,}
\newcommand{\Ryan}{R.M.L.M.\,}
\newcommand{\Gerald}{G.D.M.\,}
\newcommand{\John}{J.O.T.\,}
\newcommand{\Sarah}{S.R.D.\,}
\newcommand{\Derek}{D.F.\,}
\newcommand{\Victoria}{V.L.K.\,}
\newcommand{\Kiefl}{R.F.K.\,}
\newcommand{\Ruohong}{R.L.\,}
\newcommand{\Andrew}{W.A.M.\,}
\newcommand{\Suresh}{S.S.\,}
\newcommand{\Bob}{R.E.L.\,}

\newcommand{\SIBint}{\text{Derivation of $\BInt$}}

\makeatletter

\title{Depth-resolved Characterization of Meissner Screening Breakdown in Surface Treated Niobium}

\author[1,2]{Edward~Thoeng\thanks{Email: ethoeng@triumf.ca}}
\author[1,3]{Md~Asaduzzaman}
\author[1]{Philipp~Kolb}
\author[1]{Ryan~M.~L.~McFadden}
\author[1]{Gerald~D.~Morris}
\author[4,5]{John~O.~Ticknor}
\author[1,6]{Sarah~R.~Dunsiger}
\author[1]{Victoria~L.~Karner}
\author[1]{Derek~Fujimoto}
\author[1,3]{Tobias~Junginger}
\author[1,2,5]{Robert~F.~Kiefl}
\author[1,4,5]{W.~Andrew~MacFarlane}
\author[1]{Ruohong~Li}
\author[1]{Suresh~Saminathan}
\author[1]{Robert~E.~Laxdal\thanks{Email: lax@triumf.ca}}

\affil[1]{TRIUMF,~4004~Wesbrook~Mall,~Vancouver,~BC V6T~2A3,~Canada}
\affil[2]{Department~of~Physics~\&~Astronomy,~University~of~British~Columbia,6224~Agricultural~Road,~Vancouver,~BC V6T~1Z1,~Canada}
\affil[3]{Department~of~Physics~\&~Astronomy,~University~of~Victoria,3800~Finnerty~Road,~Victoria,~BC~V8P~5C2,~Canada}
\affil[4]{Department~of~Chemistry,~University~of~British~Columbia,~2036~Main~Mall,~Vancouver,~BC~V6T~1Z1,~Canada}
\affil[5]{Stewart~Blusson~Quantum~Matter~Institute,~University~of~British~Columbia,2355~East~Mall,~Vancouver,~BC~V6T~1Z4,~Canada}
\affil[6]{Department~of~Physics,~Simon~Fraser~University,~8888~University~Drive,~Burnaby,~BC~V5A~1S6,~Canada}

\begin{abstract}
    We report direct measurements of the magnetic field screening at the limits of the Meissner phase for two superconducting \ch{Nb} samples. 
    The samples are processed with two different surface treatments that have been developed for superconducting radio-frequency cavity applications --- a \NbBase treatment and an oxygen-doping (\NbMidTAlias) treatment. 
    The measurements show: 1) that the screening length is significantly longer in the \NbMidTAlias sample compared to the \NbBase sample; 2) that the screening length near the limits of the Meissner phase increases with applied field; 3) the evolution of the screening profile as the material transitions from the Meissner phase to the mixed phase; and 4) a demonstration of the absence of any screening profile for the highest applied field, indicative of the full flux entering the sample. 
    Measurements are performed utilizing the \gls{bnmr} technique that allows depth resolved studies of the local magnetic field within the first \SI{100}{\nano\meter} of the surface. 
    The study takes advantage of the \bSRF beamline, a new facility at TRIUMF, Canada, where field levels up to \SI{200}{\milli\tesla} are available parallel to the sample surface to replicate \gls{rf} fields near the Meissner breakdown limits of \ch{Nb}. 
\end{abstract}
    
\begin{document}
	
\flushbottom
\maketitle

\section*{Introduction}

\Gls{srf} cavities are the enabling technology for modern large-scale high-energy \gls{linac} facilities. 
The \gls{srf} cavities or resonators can efficiently produce high-amplitude \gls{rf} electromagnetic fields to accelerate charged particles~\cite{PadamseeKnoblochHays_1998, Padamsee2017,Padamsee2019}.
For a fixed final energy, the length of the linac is inversely proportional to the \gls{Ea}.
The \gls{Ea} is proportional to the peak \gls{rf} electric and magnetic fields on the cavity surface, with the surface magnetic field and associated induced screening currents limited fundamentally by the bounds of the superconducting state.
For \gls{srf} cavity applications, the superconductor must operate in the Meissner state. 
In a type-II superconductor, when the surface magnetic field exceeds the limits of the Meissner state, the material enters the mixed phase where quanta of magnetic flux circulated by vortices of supercurrents penetrate into the bulk. 
The penetrating \gls{rf} fields cause oscillation of vortices at \gls{rf} frequency that generate heat, leading to thermal instability and quench of superconductivity.
The quench field ultimately defines the maximum achievable accelerating gradient in an \gls{srf} cavity.
Achieving a higher \gls{Ea} ultimately requires a higher onset field for vortex penetration. 
Global research is focused on new processes including surface doping~\cite{Ndoping_Grassellino2013}, layered structures~\cite{Gurevich:SISpaper,Kubo:SUST}, or new materials~\cite{AnneMarie_SUST2016} that will extend the Meissner state to higher fields to support high gradient operation. 

In the Meissner state, the \gls{rf} magnetic fields are screened from the bulk within a very thin layer (\SI{\sim 100}{\nano\meter}) of the inner cavity wall of the superconducting surface.
The screening profile is typically exponential with a characteristic length given by the London penetration depth, $\lambda$~\cite{Tinkham_Book}.
For type-II superconductors, the \gls{b-c1} marks the field above which it is energetically favourable for the superconductor to be in a mixed phase though the Meissner state can be maintained in a meta-stable phase up to the \gls{b-sh} $\sim$ \gls{b-c}, made possible by the Bean-Livingston surface energy barrier~\cite{BeanLivingston1964,Kubo:SUST, LiarteTranstrum_SUST2017}. 
Niobium (\ch{Nb}), a marginal type-II superconductor~\cite{Prozorov2022_NbMarginal}, is the most common material used for \gls{srf} cavity fabrication. 
Niobium has the highest \gls{tc} among elemental (metallic) superconductors and the highest value of \gls{b-c1}  of known superconductors.

For clean niobium, \gls{b-c1}(\SI{0}{\kelvin}) is in the range of \qtyrange[range-units=single]{170}{180}{\milli\tesla}~\cite{Junginger:PRABmuSR} with \gls{b-sh}(\SI{0}{\kelvin}) estimated to be \SI{\sim 240}{\milli\tesla}~\cite{Posen_Valles_2015a}. 
The \gls{srf} community has managed to push cavity performance to peak surface RF magnetic field values beyond \gls{b-c1}~\cite{Watanabe2013,TwoStep_Grassellino2018}.
Different heat treatments have empirically been developed for processing \gls{srf} cavities --- \NbLowT~\cite{Ciovati2004}, \NbTwoStep~\cite{TwoStep_Grassellino2018}, \NbNDope~\cite{Ndoping_Grassellino2013}, and \NbMidT~\cite{He2021,Posen2015,MidT_Ito2021} --- that either reduce \gls{rf} surface resistance, increase the achievable gradient, or both. 
Material studies have shown that these treatments involve modifications in the near surface electronic mean free path as a function of distance into the material. 
Researchers are also pursuing thin films and layered structures in order to reduce the required quantity of Nb by using alternate subtrates or to increase performance.

Despite decades of success in discovering empirical solutions to improve \gls{srf} cavity performance, understanding the underlying mechanisms behind \gls{srf} dissipation is still challenging.
Conventional surface characterization techniques have been adopted, rather than tailored, for \gls{srf} studies. 
Some examples of the techniques employed are \gls{sims}~\cite{lechner_srf2021_thpfdv003, Lechner2021}, magnetization measurements~\cite{Turner2022_NSR,Casalbuoni2005}, magneto-optical imaging ~\cite{Balachandran2021_NSR}, positron annihilation spectroscopy~\cite{Wenskat2020_NSR}, and neutron radiography~\cite{Aull:JPCS2012-neutron}.
These studies have provided insight into how different recipes alter impurities at the surface and affect the superconducting properties of \gls{srf} materials.

However, the key areas of performance enhancement, namely engineering the surface layer with baking/doping or thin film coating of the \ch{Nb} surface motivate the development of a sample diagnostic, on the nanometer length scale, to characterize the local field screening.
A suitable characterization technique for \gls{srf} studies ideally would be able to measure Meissner screening/vortex penetration that fulfills three important criteria: 1) the ability to measure the local magnetic field; 2) over a nanometer-scale depth resolution; with 3) applied fields parallel to the sample surface up to \SI{\sim 200}{\milli\tesla}, to push the material near the fundamental limits of the vortex-free state. 
Non-destructive bulk measurements of the fields inside superconductors have been performed using techniques utilizing spin-polarized beams such as high-energy muons (so-called conventional muons with a nominal energy of \SI{4}{\mega\electronvolt})~\cite{Grassellino_muSR_TRIUMForig, Junginger:PRABmuSR} and neutrons~\cite{Aull:JPCS2012-neutron}.
Depth-resolved and local measurements of the surface field in \gls{srf} samples have been performed using the \gls{le-musr} technique\cite{LEmuSR_Romanenko-APL-2014,2023-McFadden-PRA-19-044018, Junginger_FEMAT2024}.
However, the \gls{le-musr} technique is limited to low applied fields of $\leq$\SI{30}{\milli\tesla}~\cite{2008-Prokscha-NIMA-595-317,2021-Prokscha-MSI-18-274} --- far less than the typical magnitude of \gls{b-c1} or \gls{b-sh} of \ch{Nb}.
Thus, the measurement of the detailed screening profile of the near surface of an \gls{srf} material at parallel fields near the limits of the Meissner state until now, has been unachievable. 

All three criteria can now be simultaneously fulfilled with the recently installed \bSRF beamline upgrade to the \gls{bnmr} facility at TRIUMF (Vancouver, Canada), which allows nanometer depth-resolved surface characterization of the local field with parallel-fields up to \SI{200}{\milli\tesla}~\cite{thoeng:bsrf_RSI}.
The \gls{bnmr} technique is complementary to \gls{le-musr}.
Instead of low energy \MuPlus, the \gls{bnmr} technique commonly uses nuclear-spin-polarized \LiEightPlus radioactive ions~\cite{macfarlane:SSNMR,macfarlane:ZPC,Morris:ISACBook}. 
The ions are significantly more massive than muons (and thus higher momentum  for the same energy) and much less deflected by the applied field that is parallel to the sample surface, but transverse to beam momentum.
The details of the instrument design, implementation, and commissioning have been reported elsewhere~\cite{thoeng:bsrf_RSI}.
Briefly, the beamline adds electrostatic optics, sample cryostat, and a Helmholtz coil to, respectively, 1) focus the \gls{rib} onto the sample, 2) cool the sample in an \gls{uhv} chamber, and 3) produce high magnetic fields up to \SI{200}{\milli\tesla} parallel to the sample surface. 
Depth control of the implanted \LiEightPlus is accomplished by adjusting the sample potential bias to vary the energy of the implanting ion beam, providing deeper penetration with higher beam energies.

We report here first experiments at \bSRF with strong parallel magnetic fields on two \ch{Nb} samples with preparations developed in the \gls{srf} community.
Our findings demonstrate: 1) a clear contrast of the magnetic penetration depth between the two \ch{Nb} samples; 2) that the screening length near the limits of the Meissner phase increases with applied field; 3) the evolution of the screening length in transitioning from Meissner to the mixed state; and 4) sample dependent surface fields marking the transition from the Meissner state to the mixed state. 
All findings demonstrate the potential of the \bSRF \textcolor{black}{beamline} to inform \gls{srf} cavity research in terms of near-surface engineering.
This new facility and the methodology showcased in this publication open a new path for future studies of novel baking/doping recipes as well as more complex layered \gls{srf} systems.

\section*{Results}

Two samples with different surface \gls{srf} processing procedures (see ‘Methods’) have been characterized at the \bSRF beamline with applied parallel fields ranging from \qtyrange[range-units=single]{100}{200}{\milli\tesla}. 
Sample 1 is prepared with a baseline treatment (\NbBase) and sample 2 is prepared with a heat treatment at \SI{400}{\degreeCelsius} for \SI{3}{\hour} commonly known in the community as \NbMidT or \NbMidTAlias. The former produces ‘clean’ \ch{Nb} with an \gls{mfp} of \SI{\sim 800}{\nano\meter} while the
heating associated with the latter surface treatment dissolves the native oxide coating of the \ch{Nb} which is then diffused into the \ch{Nb} subsurface, increasing the interstitial oxygen concentration and lowering the near surface electronic mean free path~\cite{Lechner2021, Lechner2024}.

For these measurements, both samples are \gls{zfc} to the base temperature (\SI{\sim 4.5}{\kelvin}) and, subsequently, magnetic fields parallel to the sample surface are increased monotonically in steps, with \gls{bnmr} measurements performed at each field increment.
The \gls{zfc} procedure and systematic ramping of the field steps are employed in order to remove energy barrier and geometric barrier effects~\cite{Brandt:GeometPaper} from impacting the value of initial flux penetration.
At each applied field step, depth-resolved measurements are taken at four to five different implantation energies.

The \gls{slr} technique~\cite{Slichter:Book-SLR_Ch,Mehring:PHRNMRS-1983, macfarlane:SSNMR,macfarlane:ZPC,Morris:ISACBook} is used to determine the magnetic field detected by the \LiEight probe at each depth and to quantify the field screening profile $B(x)$.
In these measurements, the applied field is parallel to the initial spin-polarization direction of the implanted \LiEight ions and polarization is lost through an energy exchange between the \LiEight probes and their surrounding environment. 
The time-dependent stochastic fluctuations of the local field that are transverse to the probe spin direction induce transitions between the \LiEight nuclear spin levels that result in “relaxation” of the initial spin state over time to thermal equilibrium (approximately zero polarization). 
The \gls{slr} rate, commonly denoted as $\RelaxRate$, characterizes the depolarization time-constant of the $\LiEight$ nuclear spins after they are stopped inside the sample.
A static superimposed field tends to slow the process of depolarization and it is the variation of $\RelaxRate$ for different probe depths that gives information on the screening profile.  

\begin{figure}[hbt!]
    \centering
    \begin{subfigure}{0.45\textwidth}
        \centering
        \includegraphics[width=\textwidth]{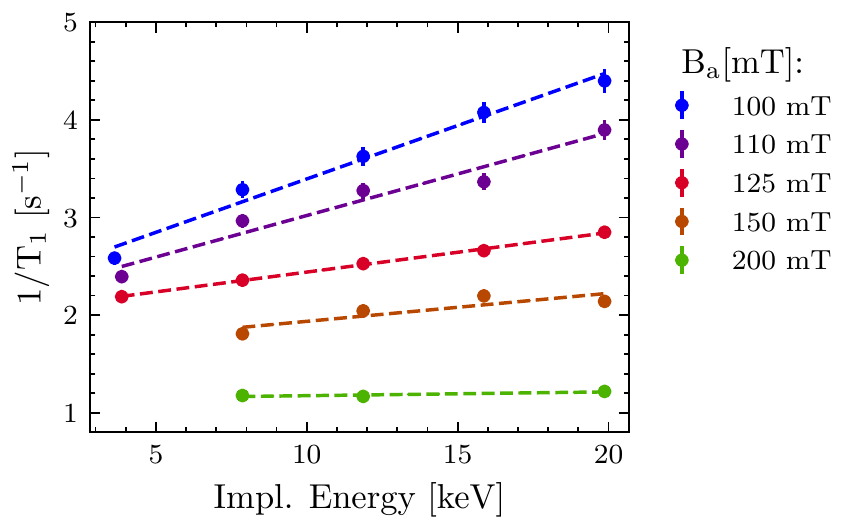}
        \caption{\NbBase sample.}
    \end{subfigure}
    \begin{subfigure}{0.45\textwidth}
        \centering
        \includegraphics[width=\textwidth]{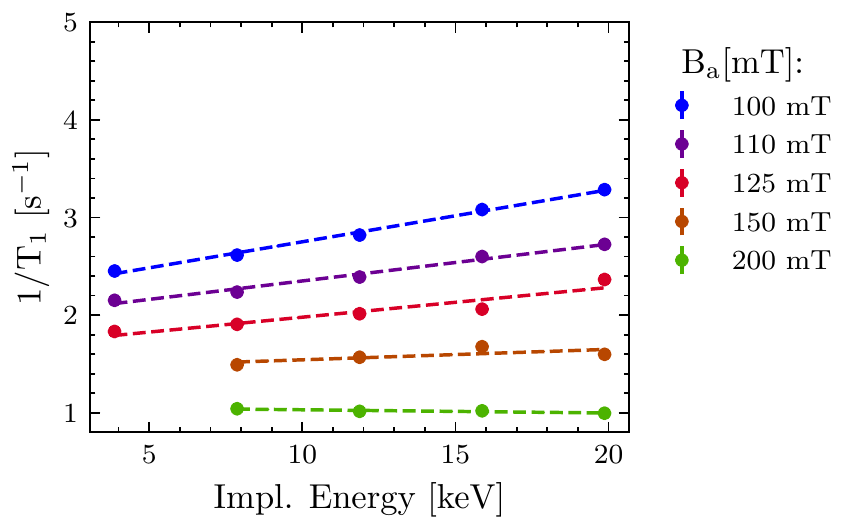}
        \caption{\NbMidTAlias sample.}
    \end{subfigure}
    \caption{\Gls{slr} rate for two \ch{Nb} samples at various applied fields, $\BApp$, as a function of implantation energy. 
    The dashed lines are simple linear fits to the measured relaxation rate to help guide the eye. The diminishing slope as $\BApp$ increases indicates the evolution from the Meissner state to the mixed state.
    }
    \label{fig:SLR_rates_vs_E}
\end{figure}

The measured \gls{slr} rates for the two samples are shown in \cref{fig:SLR_rates_vs_E} as a function of the applied field ($\BApp$) and implantation energy ($E$). 
Dashed lines in \cref{fig:SLR_rates_vs_E} are simple linear fits to the data as a guide to the eye. 
Given that $\RelaxRate\propto 1/B^2$, where $B$ is the local magnetic field for a given implantation energy,  there are several features of note.
The depolarization rate for lower applied fields is greater than at higher fields.
The rising slope of $\RelaxRate$ vs. $E$ for lower applied fields (i.e., \SI{100}{\milli\tesla}) is a clear indication of Meissner screening, as the magnetic field sampled by the ions is diminished for increasing depth.
The slope of $\RelaxRate$ decreases for higher applied fields indicating an evolution from the Meissner state to the mixed state.
Also note that the slope in relaxation rate for \SI{100}{\milli\tesla} applied field is different between the two samples indicating a surface-treatment dependence in the screening profile.
Finally, note that at \SI{200}{\milli\tesla} the $\RelaxRate$ values are independent of implantation energy (i.e., depth), indicating that the flux has substantially and uniformly entered into the sample.

A useful approximation to allow mapping $\RelaxRate$ to the average local magnetic field is through a simple Lorentzian model (see e.g., \cite{Hossain2009}): 
\begin{align}
    \frac{1}{T_1(E)} \approx \frac{a}{b + \langle B(E) \rangle^2},\label{eq:SLR_rate_simpleLor}
\end{align} 
where $E$ is the implantation energy, $\langle B(E) \rangle$ is the average measured local magnetic field, and $a$ and $b$ are material dependent fit parameters.
The $\RelaxRate$ values can be explicitly mapped to $\langle B(E) \rangle$ values by using a subset of the datapoints as a basis.
The data is taken at a constant temperature (at T\SI{\sim 4.5}{\kelvin}) to remove any temperature dependence in the Lorentzian fit parameters.
The basis points include the \SI{100}{\milli\tesla} case where the material is expected to be in the Meissner state and the \SI{200}{\milli\tesla} case where the field is assumed to be fully in the normal state near $B_{c2}$ (for $T\sim \SI{4.5}{\kelvin})$.
The analysis follows an \textit{ansatz} approach where a Meissner screening profile is assumed for \SI{100}{\milli\tesla} and the best fit to a Lorentzian is calculated by varying parameters $a$ and $b$.
A single exponential screening profile with some non-superconducting “dead layer” is assumed for the Meissner state as:
\begin{align}
    B (x) &= 
    \begin{cases}
        \BSurf &\text{for $x \leq d$,}\\
        \BSurf\exp{-(x-d)/\Lambda} &\text{for $x > d$}
    \end{cases}
    \label{eq:B_vs_x}
\end{align}
where $\BSurf$ is the field at the surface, $d$ is the “dead layer” in which there is no Meissner screening~\cite{Lindstrom2012,Lindstrom2014,Lindstrom2016}, and $\Lambda$ is the length scale parameter for the exponential decay.
 
Due to the finite volume of the sample, flux expulsion due to the superconducting diamagnetism enhances the field at the surface of the sample, $\BSurf$, with respect to the \gls{b-app} by a factor $\EnhanceFactBeamAveNoBApp$ such that $\BSurf=\EnhanceFactBeamAveNoBApp \BApp$.
For our sample geometries, the enhancement factor in the Meissner phase, $\mathcal{E}_M$, is determined using the finite element software CST Studio Suite\textregistered~\cite{CST} where the sample is assumed to be perfectly diamagnetic. 
The average value of the enhancement is determined by averaging over the Gaussian beam spot of the \LiEightPlus ions with typical size of \SI{\sim 4}{\milli\meter} in diameter~\cite{thoeng:bsrf_RSI}. 
The sample dimensions and enhancement factors for the Meissner state for each sample are summarized in 'Methods: Sample Preparations'. 
For the \SI{100}{\milli\tesla} case, $\BSurf = \mathcal{E}_M \BApp$ is assumed in the fit.
For the applied field of \SI{200}{\milli\tesla}, the diamagnetic field enhancement is expected to be negligible as evidenced by the flat $\RelaxRate$ dependence on implantation energy and so, $\mathcal{E} = 1$ is assumed in the fit.

The average field sampled by the \LiEight probes for a given implantation energy is a function of the implantation depth distribution, $\ImplDist$, and the depth profile $B(x)$.
For any particular profile, we can calculate a single value of $\langle B(E) \rangle $  as a function of energy via: 
\begin{align}
    \langle B (E)\rangle   &= \int B(x)\ImplDist dx, \label{eq:BAve}
\end{align}
where the Monte Carlo code SRIM~\cite{SRIM:Ziegler} is used to calculate the depth distributions at various beam energies (see ‘Methods: Data Analysis’).

\begin{figure}[hbt!]
    \centering
    \begin{subfigure}{0.45\textwidth}
        \centering
        \includegraphics[width=\textwidth]{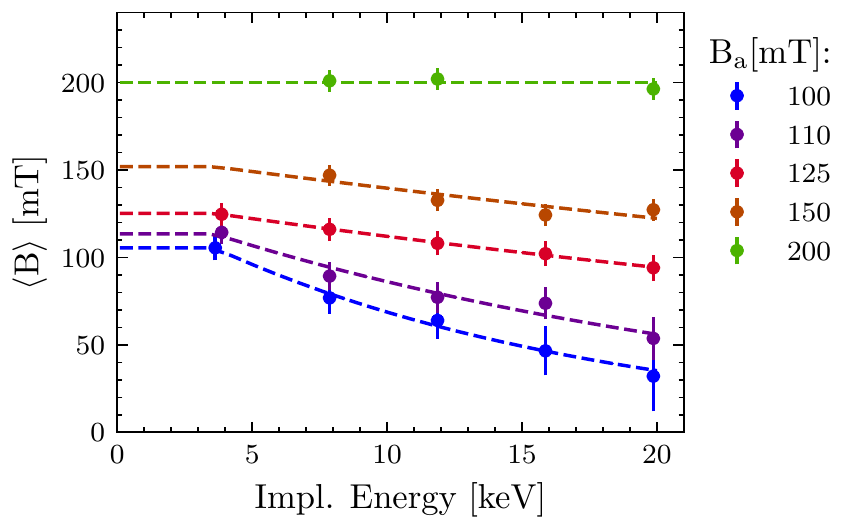}
        \caption{\NbBase sample}
    \end{subfigure}
    \begin{subfigure}{0.45\textwidth}
        \centering
        \includegraphics[width=\textwidth]{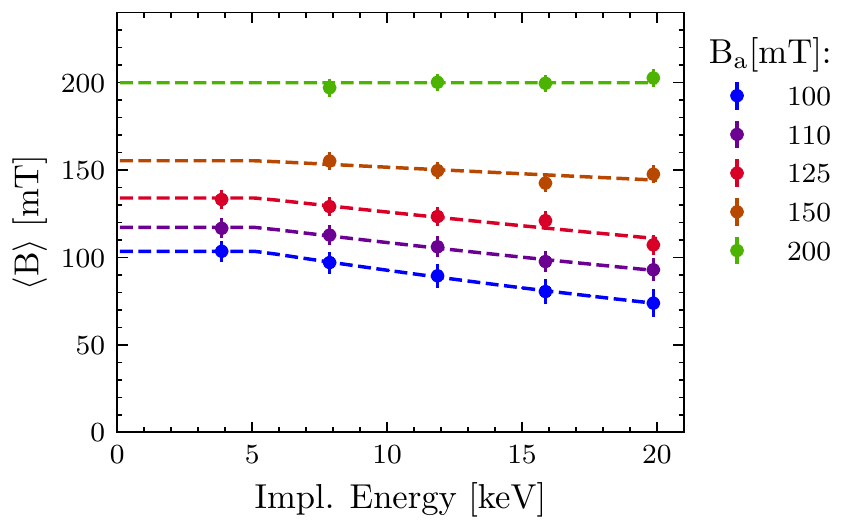}
        \caption{\NbMidTAlias sample}
    \end{subfigure}
    \caption{Shown are the average B-field as a function of implantation energy for various applied parallel fields \gls{b-app} for the two samples. Dashed lines are fits to the data assuming a single exponential function and “dead layer”.
    }
    \label{fig:BAve_vs_E}
\end{figure}

The Meissner state screening profile, defined by parameters $\Lambda$ and $d$, and the Lorentzian parameters $a$ and $b$ are varied to give the best fit of the \qtylist[list-units = single]{100;200}{\milli\tesla} $\RelaxRate$ values using the Lorentzian function in \cref{eq:SLR_rate_simpleLor}.
Assuming $\langle B(E) \rangle$ (in \unit{\milli\tesla}) and $\RelaxRate$ (in \unit{\hertz}), the best fit ($a$, $b$) parameter values are (\SI{63200(1900)}{} ,\SI{13300(1100)}{}) and (\SI{50600(1300)}{}, \SI{10000(1000)}{}) for the \NbBase and \NbMidTAlias sample, respectively.
The Lorentzian parameters are then used to map the remainder of the $\RelaxRate$ data to $\langle B(E) \rangle$.
The simple mapping results in the energy dependent screening profiles shown in \cref{fig:BAve_vs_E}.
Here, we note the evident screening at lower applied fields that systematically evolves to no screening as the applied fields increase to \SI{200}{\milli\tesla}.
Also note the distinctly different screening profiles between the two samples.

\begin{figure}[hbt!]
    \centering
    \begin{subfigure}{0.45\textwidth}
        \centering
        \includegraphics[width=\textwidth]{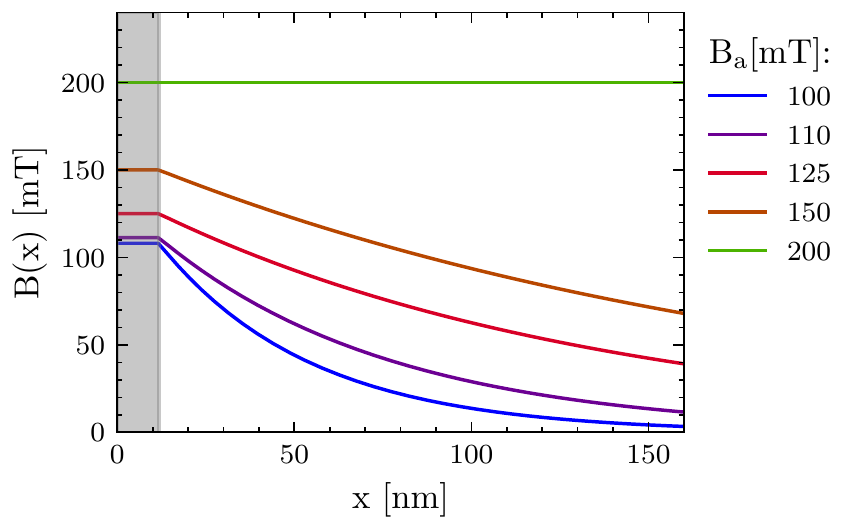}
        \caption{\NbBase sample.}
    \end{subfigure}
    \begin{subfigure}{0.45\textwidth}
        \centering
        \includegraphics[width=\textwidth]{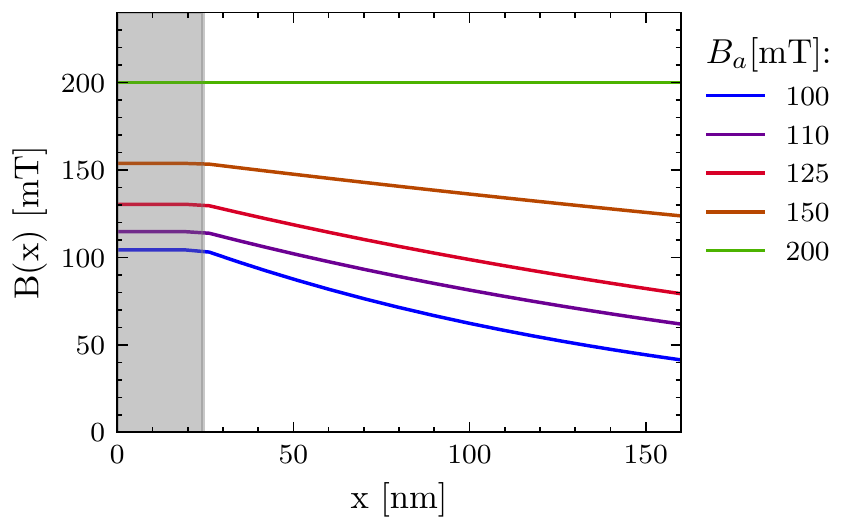}
        \caption{\NbMidTAlias sample.}
    \end{subfigure}
    \caption{
    Extracted field profiles within the first \SI{150}{\nano\meter} of the sample surface as a function of \gls{b-app}. 
    The shaded area indicates the non-superconducting “dead layer”.}
    \label{fig:B_vs_x}
\end{figure}

A more involved analysis is used to extract values of the screening length as a function of applied field.
The analysis assumes: 1) a more detailed expression for the relaxation rate which takes into account the dipole-dipole interaction between two different nuclear spins (i.e., the probe \LiEight and the host \NbNinetyThree nuclear spins)~\cite{Mehring:PHRNMRS-1983, McFadden_ThinFilmNb_2023} as described in ‘Methods: Data Analysis’; and 2) that $\RelaxRate$ values are calculated not via the average field, but via the actual field at each depth.
The latter is a more accurate technique, but it does not allow a direct mapping of $\RelaxRate$ data to an energy dependent screening profile as afforded by the analysis used to produce \cref{fig:BAve_vs_E}.

As with the previous technique, the 100 and 200 \unit{\milli\tesla} datasets (using the single exponential screening model in \cref{eq:B_vs_x} and the expected surface field enhancement factors for an applied field of 100 \unit{\milli\tesla}) are used to establish the Lorentzian-like parameters, now $B_d$ and $\tau_c$ from \cref{eq:SLR_rate_heteronuclear}.
The obtained $B_d$ and $\tau_c$ parameters are then used to extract the best fit for all datasets at various applied fields,  \gls{b-app}. 
We further introduce the normalized enhancement factor, $\normEnhanceFact(B_a)$, that relates the enhancement of the surface field compared to the applied field relative to the sample dependent maximum enhancement factor, $\mathcal{E}_M$, in the Meissner state. 
The normalized enhancement factor is defined through the relation:
\begin{align}
  \normEnhanceFact(B_a) = \frac{\EnhanceFactBeamAve-1}{\mathcal{E}_M-1}.
\end{align} 
For any superconducting volume regardless of its specific dimensions, the normalized enhancement factor should decrease from 1 when the sample is completely in the Meissner phase, to zero in the normal state.
For values in between the two limits, the volume exhibits some diamagnetic character.
In our analysis, $\normEnhanceFact=1$ is assumed for an applied field of \SI{100}{\milli\tesla} and $\normEnhanceFact=0$ for an applied field of \SI{200}{\milli\tesla}. 
For other applied fields, the data is used to extract $\normEnhanceFact$.

\begin{figure}[ht!]
\centering
    \begin{subfigure}{0.45\textwidth}
        \centering
       \includegraphics[width=\textwidth]{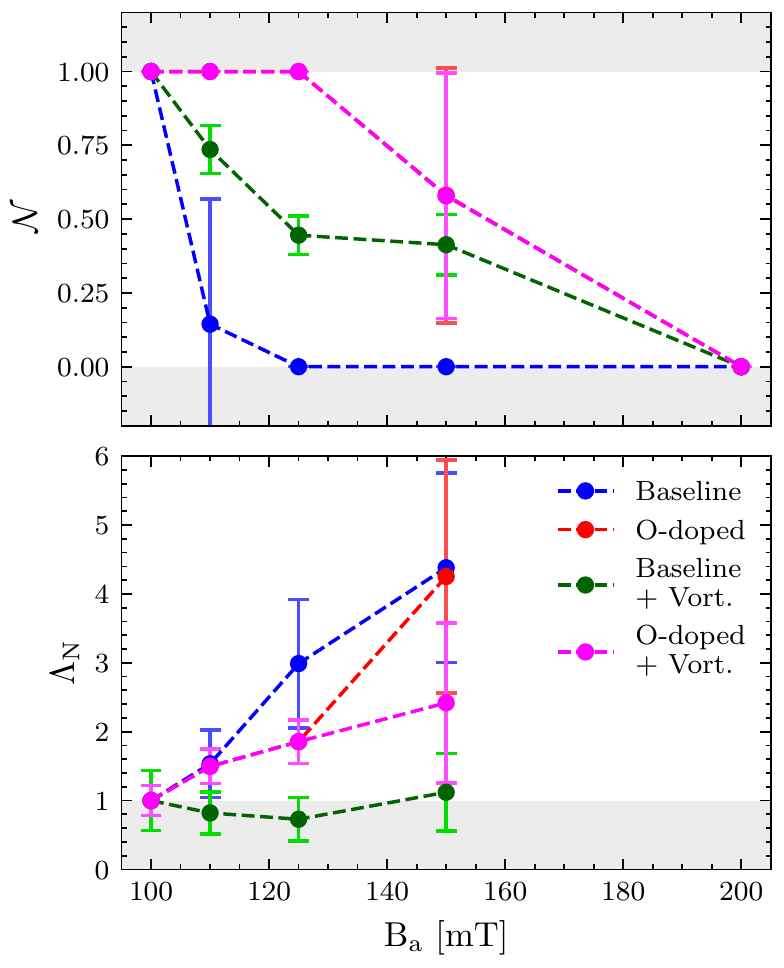}
        \caption{}
    \end{subfigure}
    \hfill
    \begin{subfigure}{0.45\textwidth}
        \centering
        \includegraphics[width=\textwidth]{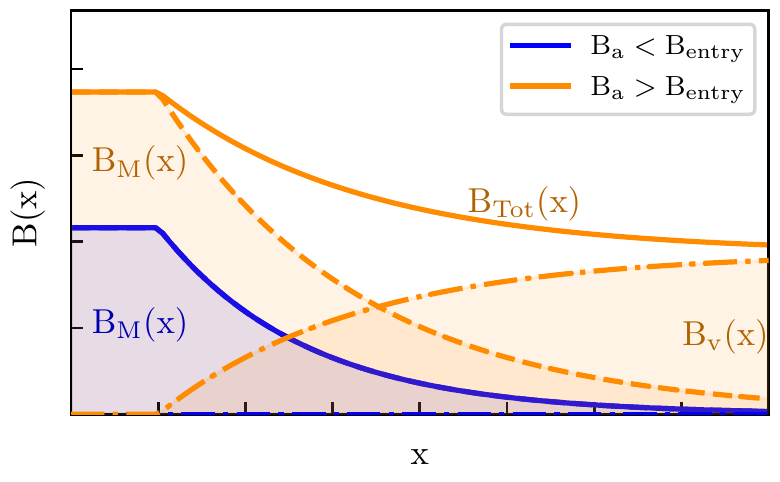}
        \caption{}
    \end{subfigure}
    \caption{
        (a) Normalized surface field enhancement factor $\mathcal{N}$ and normalized screening length $\Lambda_{N}$ for the two samples and for each analysis case (Case A, without vortex, and Case B, with vortex, as described in the text) as a function of the applied field. 
        (b) Expected screening profile from the Meissner ($\rm B_M$), vortex ($\rm B_v$), and total ($\rm B_{Tot}$) contributions for $\BApp < \BEntry$ (Meissner state) and for $\BApp > \BEntry$ with $\BInt  < \BApp $ in a mixed state with uniform vortex density. 
    } \label{fig:chi_n_normLambda}
\end{figure}

Two approaches are used in the fits. 
In the first case, Case~A, the single exponential screening profile that decays to zero, \cref{eq:B_vs_x}, is used for all applied fields.
This profile coincides with what we would expect from Meissner screening. 
In the Meissner state, the value of the normalized enhancement factor should be $\normEnhanceFact$=1 and the extracted value of $\Lambda$ corresponds to the magnetic penetration depth $\lambda$. 
However, once flux begins to break into the sample, the local field is a superposition of an exponentially decreasing field due to the screening currents and an exponentially rising field from the internal vortex distribution (see \cref{fig:chi_n_normLambda}(b)). 
The latter increases from zero at the surface towards the average bulk equilibrium value of the internal vortices near the surface~\cite{Brandt1981, Brandt1991}. 
The resulting field profile remains a single exponential, but instead of decaying to zero it reduces to the bulk equilibrium value. 
Therefore, once in the mixed state the Case~A analysis outputs an extracted $\Lambda$ that is larger than $\lambda$, and the extracted enhancement factor underestimates the actual enhancement factor.

A second analysis, Case~B, is used to better characterize the screening profile in the mixed state.  
Applied fields that yield a decrease of normalized enhancement factor $\normEnhanceFact$ from unity and an abrupt increase in screening length $\Lambda$ in Case~A are interpreted as indicating the presence of a mixed state and the screening profile is modified to~\cite{Brandt1991, Xie2022_Pinningmeissner}:
\begin{align}
    B_\text{Tot} (x) &=
    \begin{cases}
        \BSurf &\text{for $x\leq d$,} \\
        \BMeissner(x) + \BVortex(x) &\text{for $x > d$,} \\
        \qquad = \BSurf \exp{-(x-d)/\lambda}  +  \BInt\left[1 - \exp{-(x-d)/\lambda} \right]\\
        \qquad = (\BSurf-\BInt)\exp{-(x-d)/\lambda}+ \BInt,
    \end{cases}
    \label{eq:B_vs_x-vortex}
\end{align}
where $\BMeissner$ ($\BVortex$) is the Meissner (vortex) contribution to the field profile, $\BInt$ is the equilibrium field inside the bulk due to vortex entry, and the extracted characteristic length, $\lambda$, is the magnetic penetration depth. 
In order to minimize the number of fit parameters added to take into account vortex penetration, the applied field, $\BApp$, and the interior field, $\BInt$, are related through the normalized enhancement factor as $\BInt/\BApp=1-\mathcal{N}$. This simple relation is derived from the boundary condition between vacuum and superconductor layer assuming a constant effective magnetization in the volume probed by the beam (see Supplementary Information: ‘\SIBint’).

The reconstructed field profiles $B(x)$ at increasing \gls{b-app} are shown in \cref{fig:B_vs_x} for Case~A.
The extracted fit outputs for Case A and Case B are summarized in \cref{tab:Baseline-result_summary} and \cref{tab:ODoped-result_summary} for \NbBase and \NbMidTAlias samples, respectively. 
The fit outputs, as defined in \cref{eq:B_vs_x} and \cref{eq:B_vs_x-vortex}, include the screening profile length constant ($\Lambda$ in Case A, or $\lambda$ in Case B), the surface field ($B_0$), and the normalized enhancement factor ($\normEnhanceFact$), all as a function of applied field for each sample. 
We also introduce the normalized screening depth, defined as the screening characteristic depth normalized to the value at \SI{100}{\milli\tesla}, i.e., $\Lambda_N=\Lambda(B_a)/\Lambda(100$\unit{\milli\tesla}).
The plots of $\normEnhanceFact$ vs. $B_a$ and $\Lambda_N$ vs. $B_a$ are shown in \cref{fig:chi_n_normLambda}(a) for both samples and for each case. 
Note that the same values of $\normEnhanceFact(B_a)$ are obtained for both Case A and Case B for the \NbMidTAlias sample. 
At $\BApp=$\SI{150}{\milli\tesla}, despite the different $\Lambda$s obtained  for the two cases, the resulting field profiles are virtually identical.
For cases where the obtained best fit $\normEnhanceFact$ values are not at the lower (i.e., $\normEnhanceFact=0$) or upper (i.e.,  $\normEnhanceFact=1$) bound, accurate error estimates of $\normEnhanceFact$ can be obtained.  
They are shown in \cref{tab:Baseline-result_summary,tab:ODoped-result_summary} and in \cref{fig:chi_n_normLambda}(a).

\begin{table}[hbt!]
    \centering
    \begin{tabular}{|c|c|c|c|c|c|c|c|c|}
        \hline
          \multirow{2}{*}{$\BApp$ [\unit{\milli\tesla}]} & \multicolumn{4}{c|}{Case A}  & \multicolumn{4}{c|}{Case B} \\
         \cline{2-9} 
         & $\BSurf$ [\unit{\milli\tesla}] & $\normEnhanceFact(B_a)$ & $\Lambda$ [\unit{\nano\meter}] & $\Lambda_N$ & $\BSurf$ [\unit{\milli\tesla}] & $\normEnhanceFact(B_a)$ & $\lambda$ [\unit{\nano\meter}] & $\Lambda_N$ \\
         \hline
         100  & 108.1 & 1 & \SI{42.8(13.2)}{} & \SI{1.0(0.4)}{} & 108.1 & 1 &  \SI{42.8(13.2)}{} & \SI{1.0(0.4)}{} \\
         110  & \SI{111.3(3.8)}{} & \SI{0.14(0.42)}{} & \SI{65.6(5.6)}{} & \SI{1.5(0.5)}{} & \SI{116.6(0.7)}{} & \SI{0.74(0.08)}{} & \SI{35.1(7.4)}{} & \SI{0.8(0.3)}{} \\ 
         125  & 125 & \SI{0}{} & \SI{127.8(6.0)}{} & \SI{3.0(0.9)}{} & \SI{129.5(0.7)}{} & \SI{0.45(0.07)}{} & \SI{31.2(9.7)}{} & \SI{0.7(0.3)}{} \\
         150  & 150 & \SI{0}{} & \SI{187.3(10.7)}{} & \SI{4.4(1.4)}{} & \SI{155.0(1.2)}{} & \SI{0.41(0.10)}{} & \SI{47.9(19.0)}{} & \SI{1.1(0.6)}{}\\
         200  & 200 & 0 & $\infty$ & $\infty$ & 200 & 0 & $\infty$ & $\infty$ \\
         \hline
    \end{tabular}
    \caption{Extracted fit parameters from the $B(x)$ model for the \NbBase sample, measured at \SI{\sim 4.5}{\kelvin}. The thickness of the “dead layer” is estimated at $d=$\SI{11.7(2.9)}{\nano\meter}. 
    }
    \label{tab:Baseline-result_summary}
\end{table}

\begin{table}[hbt!]
    \centering
    \begin{tabular}{|c|c|c|c|c|c|c|c|}
        \hline
          \multirow{2}{*}{$\BApp$ [\unit{\milli\tesla}]} & \multirow{2}{*}{$\BSurf$ [\unit{\milli\tesla}]} & \multirow{2}{*}{$\normEnhanceFact(B_a)$}  &  \multicolumn{2}{c|}{Case A}  & \multicolumn{2}{c|}{Case B} \\
         \cline{4-7} 
           &  &  & $\Lambda$ [\unit{\nano\meter}] & $\Lambda_N$ & $\lambda$ [\unit{\nano\meter}] & $\Lambda_N$ \\
         \hline
         100  & 104.3 & 1 & \SI{146.9(22.6)}{} & \SI{1.0(0.2)}{}  &  \SI{146.9(22.6)}{} & \SI{1.0(0.2)}{} \\
         110  & 114.7 & 1 & \SI{219.8(14.3)}{} & \SI{1.5(0.3)}{}  & \SI{219.8(14.3)}{} & \SI{1.5(0.3)}{} \\
         125  & 130.4 & 1 &  \SI{272.5(19.4)}{} & \SI{1.9(0.3)}{}  & \SI{272.5(19.4)}{} & \SI{1.9(0.3)}{} \\
         150  & \SI{153.8(2.8)}{} & \SI{0.58(0.42)}{} & \SI{624.7(229.0)}{} & \SI{4.3(1.7)}{}  & \SI{355.2(161.4)}{} & \SI{2.4(1.2)}{}\\
         200  & 200 & 0 & $\infty$ & $\infty$  & $\infty$ & $\infty$\\
         \hline
    \end{tabular}
    \caption{Extracted fit parameters from the $B(x)$ model for the \NbMidTAlias sample, measured at \SI{\sim 4.5}{\kelvin}. 
    The thickness of the “dead layer” is estimated at $d=$\SI{24.4(11.2)}{\nano\meter}. 
    }
    \label{tab:ODoped-result_summary}
\end{table}

\section*{Discussion}

The results reveal several notable features. 
They are: 1) the screening length in the \NbMidTAlias sample is $\num{\sim 3}$ times longer than in the \NbBase sample; 2) the screening length in the \NbMidTAlias sample increases significantly as the applied field is increased; 3) flux-penetration into the bulk occurs at a lower field in the \NbBase treatment than in the \NbMidTAlias sample; and 4) the \NbMidTAlias sample possesses a larger non-superconducting “dead layer” than the \NbBase sample.

First, the \NbMidTAlias sample has a significantly larger penetration depth compared to the \NbBase.
This is consistent with the expectation that mid-T baking (\SI{400}{\degreeCelsius} for \SI{3}{\hour}) renders a “dirtier” (shorter mean free path) subsurface due to native oxide (\ch{Nb2O5}) dissolution and interstitial oxygen diffusion~\cite{Ciovati2004,Lechner2021, Lechner2024}.
In contrast, the \NbBase sample has a much shorter penetration depth in the Meissner phase consistent with a cleaner \ch{Nb} surface.

In type-II superconductors, the mixed state transitions to the normal state and the superconducting screening collapses at $\BcTwo(t)$ with temperature dependence~\cite{Tinkham1963_Bc2Formula,Finnemore1966}:
\begin{align}
    b_{c2} = \frac{1-t^2}{1+t^2},\label{eq:Bc2_Tdep_reduced}
\end{align}
where $t=T/T_c(B=0)$ is the reduced temperature and $b_{c2}=\BcTwo(T)/\BcTwo(T=0)$ is the reduced upper critical field at temperature $T$.
The relation in \cref{eq:Bc2_Tdep_reduced} can be inverted to give the reduced critical temperature at a reduced field $b=\BSurf/\BcTwo(T=0)$:
\begin{align}
    t_{c2} &= \sqrt{\frac{1-b}{1+b}}\label{eq:Tc2_Bdep_reduced},
\end{align}
where $t_{c2} = T_{c2}(\BSurf)/T_{c2}(B=0)$.

The two-fluid model can be used to estimate the effective penetration depth, $\lambda$(0,0), at $T=0$ K and $B\rightarrow 0$  via:
\begin{align}
   \lambda (0,0) &= \mpd(t,b) \cdot {\sqrt{1-[t(b)]^4}}\label{eq:mpd_T_B_dep},
\end{align}
where the field-dependence of the penetration depth arises due to the field dependence of $T_{c2}(\BSurf)$ with the reduced temperature defined earlier as $t(b) = T/T_{c2}(\BSurf)$.
In the local limit, the relation between $\mpdZeroZero$ and the \acrfull{mfp} can be estimated from the relation~\cite{Tinkham_Book}:
\begin{align}
    \mpdZeroZero = \lpdZero \sqrt{1 + \frac{\xiZero}{\mfp}}\label{eq:mpd_mfp_dep}.
\end{align}
where $\lambda_L$ is defined as the London penetration depth
in the limit of pure ($\mfp \rightarrow \infty$) and local ($\xiZero \rightarrow 0$) response and $\xiZero$ is the \gls{bcs} coherence length~\cite{BCS_1957}.

Typical quoted values for clean niobium are $T_{c2}(B=0)$ \SI{\sim 9.2}{\kelvin}~\cite{Finnemore1966, Casalbuoni2005}, $\BcTwo(T=0)$ \SI{\sim 410}{\milli\tesla}~\cite{Finnemore1966, Casalbuoni2005}, $\lambda_L$ \SI{\sim 40}{nm} ~\cite{Gurevich2017, Maxfield1965_PenDepthRef} and $\xiZero$=\SI{38}{nm}~\cite{Maxfield1965_PenDepthRef}. 
Our \NbBase sample is characterized by a \gls{RRR}=300, corresponding to a mean free path of \gls{mfp} \SI{\sim 810}{\nano\meter} from the empirical relation: $\mfp\text{[nm]} \approxeq 2.7 \cdot \text{RRR}$~\cite{PadamseeKnoblochHays_1998}. 
Applying \cref{eq:Bc2_Tdep_reduced,eq:Tc2_Bdep_reduced,eq:mpd_T_B_dep,eq:mpd_mfp_dep} for our \NbBase sample with measured values of $\mpd(T=\SI{4.5}{\kelvin}, \BSurf=\SI{108.1}{\milli\tesla}, \mfp=\SI{810}{\nano\meter})=$ \SI{42.8}{\nano\meter} gives $\mpd(0, 0, \mfp=810) =$ \SI{39.0}{\nano\meter} or  $\lpd$=\SI{38.1}{\nano\meter} in good agreement with expected values.  

The same analysis can also be applied for the \NbMidTAlias sample by using the same input values of $T_c=$\SI{9.2}{\kelvin}, $B_{c2}=$\SI{410}{\milli\tesla} and $\lpd=$\SI{38.1}{\nano\meter}
The experimental value of $\mpd(t,b)=$\SI{146.9}{\nano\meter} gives a penetration depth of $\mpd(0,0)=$\SI{134.5}{\nano\meter} and a mean free path value of $\mfp=$\SI{3.3}{\nano\meter} from \cref{eq:mpd_mfp_dep}.
Mean free path values of doped \ch{Nb} have been reported by other studies from fits to RF cavity measurements with values in the range of \qtyrange[range-units=single]{4}{200}{\nano\meter}~\cite{Maniscalco2017,koufalis2017_mfp}. 
The extracted values here are on the low side of this range, which might indicate additional pollution in the surface layer and/or suppressed \gls{tc} due to an elevated concentration of oxygen impurities~\cite{Koch1974,DeSorbo1963}.  
This latter hypothesis can be used to also look at a second key feature of the data from the \NbMidTAlias sample, the increase in screening length with applied field. 

We consider first the break-in field implied by the data. 
Considering the sample geometry and formulation by Brandt for a sample with rectangular cross-section with side length $c$ in the direction of the applied field and side $a$ as the slab thickness, flux will break into the center of a pin free sample at an applied field of~\cite{Brandt:GeometPaper}: 
\begin{align}
    \BEntry = \BcOne\times \tanh(\sqrt{0.36 c/a}).\label{eq:B-entry-Brandt}    
\end{align}
Using the experimentally measured value of $B_{c1}(0 K)$\SI{\approx 174}{\milli\tesla}~\cite{Junginger:PRABmuSR} and scaling this value to the measurement temperature of \SI{\sim 4.5}{\kelvin} using $B_{c1}(T)=B_{c1}(0) [1-(T/T_c)^2]$~\cite{Finnemore1966}
gives $B_{c1}$(\SI{4.5}{\kelvin})\SI{\approx 132}{\milli\tesla}.
Noting that the samples are slightly different in shape (see \cref{tab:sample_dim}) and applying \cref{eq:B-entry-Brandt}, the entry field would be expected at an applied field of $\BEntry$\SI{\sim 113}{\milli\tesla} in the \NbBase and $\BEntry$\SI{\sim 123}{\milli\tesla} in the \NbMidTAlias sample. 
In the dirtier \NbMidTAlias sample, we expect that the assumed $\BcOne$ value is reduced from the clean limit value and so this $\BEntry$ value represents an upper bound.
There is indication from the normalized enhancement factor fits that flux has broken into the \NbBase sample already at an applied field of \SI{110}{\milli\tesla} ($\normEnhanceFact<1$), but in the \NbMidTAlias case the extracted enhancement factor is consistent with the Meissner state ($\normEnhanceFact=1$) for applied fields from \qtyrange[range-units=single]{100}{125}{\milli\tesla} (see \cref{fig:chi_n_normLambda} (a)). 

\begin{figure}[hbt!]
    \centering
    \includegraphics[width=0.5\textwidth]{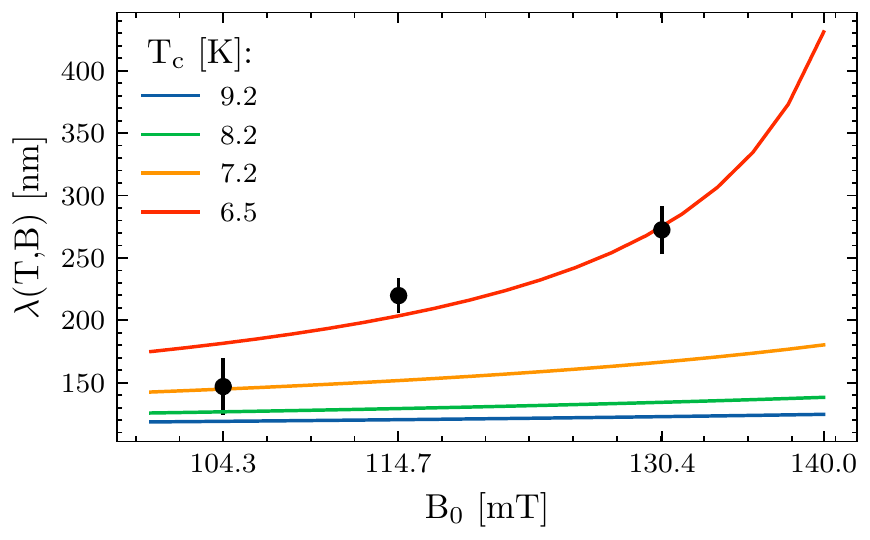}
    \caption{Strong field-dependence of the measured effective magnetic penetration depth (data points) for the \NbMidTAlias sample, compared to a linear Meissner penetration depth model for various reduced critical temperatures $T_c$.
    Note that the penetration depth values are plotted against the surface fields, i.e., $\BSurf = \EnhanceFactBeamAveNoBApp \BApp$.
    } 
    \label{fig:screenlen_vs_BSurf_MidT}
\end{figure}

For the \NbMidTAlias cases, the measured magnetic penetration depth $\lambda$ increases significantly as the surface field nears the boundary of the Meissner state, as shown in \cref{fig:chi_n_normLambda}(a).
A way to account for this sharp rise is to consider a suppressed \gls{tc}.
A decrease of \gls{tc} for \ch{Nb} has been reported due to interstitial oxygen diffusion in \ch{Nb}~\cite{Koch1974,DeSorbo1963}. 
A linear fit to the tabulated values of \gls{tc} vs \ch{O} concentration from Koch et al.~\cite{Koch1974} is approximately given by $T_c \approxeq T_{c0} - 1.01\cdot x$, where $T_{c0}=$\SI{9.2}{\kelvin} is the \gls{tc} of pure \ch{Nb} and $x$ is the atomic-\% concentration of interstitial oxygen in \ch{Nb}\cite{Koch1974}.
From \cref{eq:mpd_T_B_dep,eq:Tc2_Bdep_reduced}, and using \gls{tc} as the free variable with fixed $B_{c2}=\SI{410}{\milli\tesla}$, a best fit to the three points in ~\cref{fig:screenlen_vs_BSurf_MidT} is for a suppressed \gls{tc} \SI{\sim 6.5}{\kelvin}, with $\lambda(0,0)$\SI{\sim 103}{\nano\meter} and \gls{mfp} \SI{\sim 6}{\nano\meter} obtained from \cref{eq:mpd_mfp_dep} by using the same values of $\lpd$ and $\xiZero$ as inputs. 
This level of suppression would require an oxygen concentration of a few atomic-\%, much higher than typical values quoted for \NbMidTAlias treatments that are $\sim 1/10$-th of this concentration ($\lesssim  0.5$ [at\%])~\cite{Lechner2021,lechner_srf2021_thpfdv003}. 
Alternatively, the strong change in screening profile, assuming $T_{c0}=$\SI{9.2}{\kelvin}, could also be fit by a suppressed upper critical field with $B_{c2}(\SI{0}{\kelvin})\sim\SI{230}{\milli\tesla}$, or a combination of suppressed $T_c$ and suppressed $B_{c2}$.
The simultaneous suppression of $T_c$ and $B_{c2}$ can be due to the presence of pair-breaking magnetic impurities (see, e.g., \cite{Kogan2013,Kogan2014,Kogan2022}) as observed in experimental measurements (see, e.g., \cite{Bawa_NSR2016}), and is consistent with calculations from the microscopic Eilenberger's equations~\cite{Eilenberger1968}.

Non-magnetic impurities in current carrying superconductors act as pair breakers, in contrast to their pair conserving nature in zero current, thereby reducing the pair potential $\Delta$ and superfluid density~\cite{Lin_Gurevich_2012, Gurevich2017,Gurevich2023, Kubo_PRA2022}.
The dependence of superfluid density on screening currents in the Meissner state contributes to non-linear Meissner screening  as $\Delta^2(\BSurf) \propto n_s(\BSurf)\propto$1/$\lambda^2(\BSurf)$~\cite{Gurevich2017,Gurevich2023}.
The non-linear screening adds a field-dependent correction term to the magnetic penetration depth on top of implicit $\BSurf$ dependence via $T_{c2}(\BSurf)$ in \cref{eq:mpd_T_B_dep}. 
For s-wave superconductors it is predicted that the penetration depth increases quadratically with field~\cite{Yip1992,Xu1995}:
\begin{align}
    \mpd_\text{NLME}(t,b) &= \mpd(t,b)\left[ 1 + \beta(T) \left(\frac{\BSurf}{B_c}\right)^2\right]\label{eq:mpd_NLME}.
\end{align}
Here, $\mpd(t,b)$, $t$ and $b$ are  defined in \cref{eq:mpd_T_B_dep} and $\beta(T)$ is the quadratic prefactor. 
Ginzburg-Landau theory provides an analytical expression for the prefactor $\beta(T\rightarrow T_c)$, which depends only on the Ginzburg-Landau parameter $\kappaGL$ (defined as $\lambda/\xi$ for $T\rightarrow T_c$) via the following relation~\cite{Makita_Gurevich2022,Ginzburg1950}:
\begin{align}
    \beta &= \frac{\kappaGL(\kappaGL + 2^{3/2})}{8(\kappaGL + 2^{1/2})^2}.\label{eq:beta_NLME_GL}
\end{align}
Typical $\kappaGL$ values of $1\rightarrow 10$ (i.e., in the dirty limit $\kappa\approx\lambda_L/\mfp$) 
for various levels of doping in niobium give $\beta\sim$0.1. For the range of values reported here, with $B_c\sim$\SI{200}{\milli\tesla},  the non-linear Meissner screening should enhance the magnetic penetration depth at \SI{125}{\milli\tesla} only by <2\% above the value at \SI{100}{\milli\tesla}. Our reported difference of almost a factor of two is considerably larger than that.

More extreme variations of the screening profile near \gls{b-c1} could come from additional pair-breaking from magnetic impurities or gap anisotropy, resulting in a subgap state that is phenomenologically modeled by density of states with Dynes broadening~\cite{Kubo_PRA2022, Kubo_SUST2021, Gurevich2023}. 
For example, magnetic impurities associated with oxygen vacancies in the native surface oxide of Nb have been revealed by tunneling measurements and have been attributed as an intrinsic component to the residual RF resistance commonly encountered in SRF cavities~\cite{Proslier_IEEE2011,KharitonovProslier_YSR_PRB2012, 2017-Junginger-SST-30-125012, Wenskat2022_surfmag}. 
These subgap states could result in a stronger non-linear field dependence of the Meissner screening similar to the “extreme non-linear Meissner effect” in the gapless state as reported by Lee et al~\cite{Lee2023}.

Another possible mechanism to partially explain the strong dependence of magnetic penetration depth with applied field near \gls{b-c1} is localized nucleation of vortex loops at the surface~\cite{Genenko1998,asad2024_muSR}.
Given the doping used in the \NbMidTAlias sample, it would be expected to have an elevated $\kappa$ and a reduced \gls{b-c1}. 
However, given the observed surface field enhancement (i.e., $\normEnhanceFact=1$) which indicates that the bulk of the material remains diamagnetic (see \cref{fig:chi_n_normLambda}(a)), it is likely that the material is operating in a regime above $B_{c1}$ in a meta-stable regime below $B_{sh}$. 
In this regime, variations in the order parameter from local suppression of $T_c$, impurities, inclusions, or field enhancement from surface roughness  could suppress locally the superheating barrier and be responsible for the surface nucleation of vortex loops. 
Such vortices would not penetrate into the bulk, but would increase in number as the applied field is increased, right up to the applied field where flux penetration occurs. 
These vortex loops at the surface could contribute to  the large change in effective penetration depth measured in the doped sample due to the superposition of the Meissner screening profile and the additional surface field from nucleating surface vortices. 

Our results in Case B reveal interesting features of the mixed state. As noted in \cref{fig:chi_n_normLambda}(b), after vortices have entered the bulk, the resulting screening profile is a superposition of the exponentially decaying field due to the Meissner screening and the exponentially rising contribution due to the vortex entry that is zero at the surface and rises to the near surface vortex density. 
While our simple model assumes a uniform flux distribution, the details of the vortex density are defined by the sample geometry~\cite{Brandt:GeometPaper}, the average field in the bulk, the local pinning force~\cite{Xie2022_Pinningmeissner,asad2024_muSR}, as well as additional field dependence of penetration depth $\mpd(\BApp)$~\cite{Lin_Gurevich_2012, Kubo2020,Kubo_PRA2022}.
\Cref{fig:chi_n_normLambda}(a) for Case B indicates that the extracted magnetic penetration depth for the \NbBase sample has a reduced field dependence compared to the \NbMidTAlias sample, consistent with “clean” niobium and a pairing potential that depends only weakly on the applied field. 
Another interpretation is that some flux may have penetrated the \NbMidTAlias sample giving a non-zero $\BInt$ (\cref{eq:B_vs_x-vortex}), but due to flux pinning the vortices do not redistribute uniformly and remain near the surface after entry.
Such an interpretation could be consistent with an increasing screening profile as a function of applied field, but with a surface field enhancement factor $\normEnhanceFact\sim 1$. 
Subsequent studies could be done to evaluate any hysteresis in the screening profile. 
In general, future experiments could explore near surface magnetic field in the mixed state with more detailed measurements in this field regime. 

A notable feature in the data is the shorter “dead layer” in the \NbBase sample ($d$ = \SI{11.7(2.9)}{\nano\meter}) compared to the \NbMidTAlias sample ($d$ = \SI{24.4(11.2)}{\nano\meter}).
The “dead layer” is a universal feature of all superconductors and is a sample dependent property.
The origin of the “dead layer” has previously been attributed to surface roughness (see e.g.\cite{Lindstrom2012,Lindstrom2014,Lindstrom2016}).
As both samples are from the same batch with the same \gls{bcp} etching step applied, it is unlikely that surface roughness variation is the sole origin of the much larger “dead layer” in the \NbMidTAlias sample.
Surface composition characterization has been performed on duplicate witness samples with equivalent treatment using \gls{tof-sims} and \gls{edx}, and has been reported by Kolb et al~\cite{kolb2023midt}.
Both characterizations indicate an elevated level of carbon concentration at the surface of the \NbMidTAlias sample, possibly received during the \SI{400}{\degreeCelsius}-bake. 
The presence of enhanced carbon concentration at the surface is consistent with other studies, e.g., in-situ \gls{xps}, where furnace baking of \ch{Nb} at \SI{400}{\celsius} for \SI{3}{\hour} dissolves the protective \ch{Nb2O5} layer below the critical thickness of \SI{1}{\nano\meter}, thereby promoting surface carbon precipitation~\cite{prudnikava2024insitu}.
Strong impurity gradients over the first \SI{20}{nm} have been postulated to modify the magnetic screening profile to non-exponential forms and push peak screening currents away from the surface \cite{Checchin2020_diffProfile, Ngampruetikorn_Sauls_2019,McFadden2024_Comments, Lechner2024}.  
Such modified screening profiles with a dead-layer magnitude similar to the \NbBase sample (i.e., $d$\SI{\sim 12}{nm}) may be consistent with our data that we have modeled simply as a single exponential with a larger “dead layer” and enhanced penetration depth. 
It is also important to note that pair-breaking due to screening currents will vary over the penetration depth $\lambda$.
This pair-breaking modifies the super-fluid density locally as a function of the screening depth, i.e., $\Delta[B(x)]^2 \propto n_s(x)\propto$1/$\lambda(x)^2$, and therefore results in a non-exponential magnetic screening profile at high field~\cite{Kubo2020}.

In summary, the performance of \gls{srf} materials near fundamental limits is of critical importance to the accelerator community, as these limits set the scale for the accelerators in terms of operating gradient. 
Relevant phenomena in this regime have been explored in detail  using two heat treated \ch{Nb} samples, both showing distinct behaviour as a result of their modified subsurface.
Theories which self-consistently take into account various pair breaking effects (due to magnetic field, nonmagnetic and magnetic impurities, gap anisotropy, proximity effect from normal metal overlayer) are being extensively developed (see, e.g.,\cite{Kubo_PRA2022,Kubo_SUST2021,Gurevich_Kubo_2017a,Gurevich_Kubo_2019,Ngampruetikorn_Sauls_2019,Gurevich2023,Zarea2023_DFT}).
We anticipate that the \bSRF facility will be an important tool to validate these proposed theories, and future experiments will help refine our understanding on how to advance and tailor \gls{srf} performance.

\section*{Methods}

\subsection*{Sample Preparations\label{sec:sample_prep}}

The sample size is constrained to roughly \qtyproduct{10 x 10 x 2}{\milli\meter} by the mechanical dimensions of the sample ladder in the \bSRF cryostat~\cite{thoeng:bsrf_RSI}.
The \ch{Nb} samples are cutouts of \gls{RRR} $\gtrsim 300$ \ch{Nb} sheets sourced from ATI Wah Chang (Albany, Oregon, USA).
The samples are prepared simultaneously during processing of bulk \ch{Nb} \gls{rf} cavities. 
All \ch{Nb} samples first undergo a \NbBase treatment as follows:
\begin{enumerate}

    \item \Gls{bcp}, a standard chemical etching solution for \gls{srf} cavities containing 2:1:1 volume ratio of \ch{H3PO4}:\ch{HNO3}:\ch{HF}, to remove the first \SI{50}{\micro\meter} of damaged layers due to cutting/machining.
    Studies of \gls{bcp}-treated niobium report typical roughness values of \SI{\sim 1.6}{\micro\meter}~\cite{Xu2011_BCPRoughness}.
    
    \item Annealing in-house with a high-vacuum furnace at \SI{1400}{\degreeCelsius} for \SI{\sim 4}{\hour} in order to remove pinning sources in the material~\cite{Junginger:PRABmuSR}.
    During this annealing step, all the samples are wrapped in a clean \ch{Nb} foil. 
    
    \item An additional \SI{5}{\micro\meter} \gls{bcp} is applied to remove possible furnace contaminations.
    
    \item Ultrasound-cleaning with de-ionized water.
    
\end{enumerate}

\begin{figure}[hbt!]
    \centering
    \includegraphics[width=0.6\linewidth]{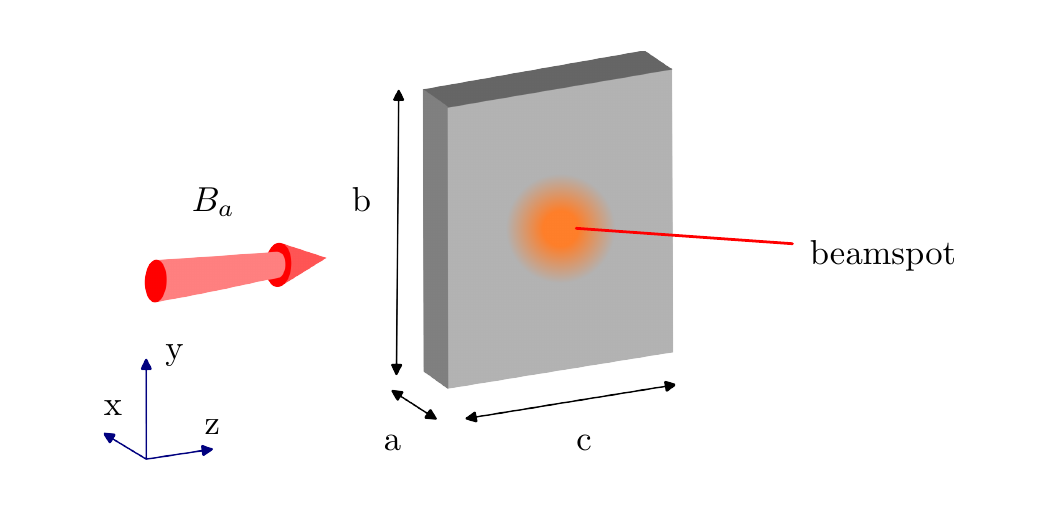}
    \caption{
        Sample dimensions and orientation with respect to the applied magnetic field and beam implantation. The beamspot is roughly to scale. 
    } \label{fig:CST_field}
\end{figure}

A second sample was additionally treated with a \NbMidT or \NbMidTAlias recipe where the cavity/sample receives a “bake” of \SI{400}{\degreeCelsius} for \SI{3}{\hour} under ultra-high vacuum.
The dimensions of both \NbBase and \NbMidTAlias samples are shown in \cref{tab:sample_dim} with axis parameters, approximate beam spot size, and applied field orientation defined in \cref{fig:CST_field}.
Also listed are $\mathcal{E}_M$, the Meissner state surface field enhancement factor, and  $\BEntry$, the expected applied field where flux would break into the sample center, assuming $\BcOne$=\SI{174}{mT} at \SI{0}{K}.

\begin{table}[hbt!]
    \centering
    \begin{tabular}{|c | c | c|}
        \hline
        Parameter & \NbBase & \NbMidTAlias \\
        \hline
        a (\unit{\milli\meter}) & 1.8 & 1.7  \\
        b (\unit{\milli\meter}) & 9.8 & 12.4  \\
        c (\unit{\milli\meter}) & 8.1 & 12.6  \\
       $\mathcal{E}_M$ & 1.081 & 1.043 \\
        \BEntry (\unit{\milli\tesla}) @ \SI{4.5}{\kelvin} & 113 & 123  \\
        \hline
    \end{tabular}
    \caption{Sample dimensions and the calculated geometrical enhancement factor $\mathcal{E}_M$ in the Meissner state from finite element simulations and the applied field, \BEntry, where entry into the bulk is expected. 
    \label{tab:sample_dim}}
\end{table}

\subsection*{\bNMR Experiment of \Gls{srf} Samples}\label{sec:bNMR_exp}

The \bNMR technique~\cite{macfarlane:SSNMR,macfarlane:ZPC} implants spin-polarized radioactive ions and measures the local magnetic field at the stopping site of the probe via detection of the emitted $\beta$-particles.
The emission of the $\beta$-particles is highly anisotropic and correlated with the direction of the probe's nuclear spin at the time of the decay.
By monitoring the counts of the $\beta$-particles, the local magnetic field can be measured via its influence in reorienting the nuclear spin of the probe.
Importantly, \bNMR provides spatial resolution on the nanometer scale, achieved through the control of the probe's implantation energy.

The principle of \bNMR is similar to \gls{le-musr}~\cite{2008-Prokscha-NIMA-595-317,2021-Prokscha-MSI-18-274}, but differs in the properties of the probe in use (e.g., radioactive lifetime $\tau$\SI{\sim 1.2}{\second} for \LiEight vs $\tau$\SI{\sim 2.2}{\micro\second} for \MuPlus). 
This difference makes the two techniques complementary, as well as allowing different measurements of phenomena occuring over longer timescales. 
The radioactive ions used in \bNMR need to be spin-polarized in-flight (via optical pumping), whereas \MuPlus are intrinsically spin-polarized upon production.
A dedicated facility with \gls{rib} production, laser polarization, and beam transport is needed before the beam can be delivered to the spectrometer for depth-resolved measurements.
The following discussion provides more details on the \gls{rib} production, polarization, delivery, and measurements of the \bNMR facility.

\subsubsection*{Spin-polarized Probe Production \& Beam Transport}

High intensity \glspl{rib} are routinely produced at the TRIUMF \gls{isac} facility~\cite{Dilling:ISACBook} using high-intensity protons from the \SI{500}{\mega\electronvolt} cyclotron as a driver to bombard a solid target. 
The \bNMR radioisotope nuclei \LiEight (nuclear spin $I=2$; gyromagnetic ratio $\gLi/2\pi$ = 6.30198(8) MHz/T~\cite{2019-Stone-INDC-NDS-0794}, electric quadrupole moment $Q$ = \SI{+31.4}{\bohrmagneton}~\cite{2021-Stone-INDC-NDS-0833}; radioactive lifetime $\tau_\beta$ = \SI{\sim 1.21}{\second}~\cite{2010-Flechard-PRC-82-027309}; and mass $A_{^{8}Li}$ = \SI{8.02}{\amu}) are produced with the \gls{isol} method where they are ionized, extracted, and separated on-line to produce isotopically pure \glspl{rib} for various experiments.
At the TRIUMF \gls{bnmr} facility, low-energy (\qtyrange[range-units=single]{20}{30}{\kilo\electronvolt}) \LiEightPlus ions of \SI{\sim e7}{\per\second} are routinely delivered for experiments~\cite{macfarlane:SSNMR,macfarlane:ZPC,Morris:ISACBook}.
During delivery to the experiment, the ions are neutralized and spin-polarized in-flight at the polarizer facility via collinear optical pumping with circularly polarized resonant laser light~\cite{Levy:ISACBook,2023-Li-NIMB-541-228}, yielding a high degree (\SI{\sim 70}{\percent}\cite{MacFarlane2014_InitSpinPol}) of nuclear spin-polarization.
They are afterwards re-ionized before delivery to the sample location.

\subsubsection*{High-parallel-field Spectrometer: “\bSRF beamline”}

The spin-polarized ions can be sent for experiments on either one of two existing beamlines: the \gls{bnmr} beamline with up to \SI{9}{\tesla} applied fields (perpendicular to sample surface \& parallel to beam momentum), and the \bNQR beamline with one spectrometer station capable of fields up to \SI{24}{\milli\tesla} and a second new spectrometer, highlighted here, capable of \SI{200}{\milli\tesla} applied fields (each with applied field parallel to the sample surface \& transverse to the beam momentum).
Each spectrometer is equipped with a magnet coil outside the \gls{uhv} chamber, a pair of scintillation $\beta$ detectors, an \gls{rf} coil inside the \gls{uhv} chamber, and a cold-finger cryostat. The cryostat allows sample cooling down to \SI{\sim 4}{\kelvin} and is mounted on an electrically isolated \gls{hv} platform to allow beam energy deceleration within a short distance from the sample for depth-resolved measurements.

Studies of \gls{srf} materials require fields parallel to the sample surface to emulate the \gls{rf} magnetic field direction with respect to the \gls{srf} cavity wall and to provide a uniform surface field over the beamspot.
This is challenging to achieve with a low-energy ion beam, due to the strong bending of the beam's trajectory in this field geometry. To overcome this, a novel beamline extension has been designed and implemented~\cite{thoeng:bsrf_RSI}, allowing for field strengths of relevance for \gls{srf} materials to be employed.

\subsection*{Data Analysis\label{sec:data_analysis}}

The local magnetic field profile can be obtained via depth-resolved measurements of the \LiEight \gls{slr} rate ~\cite{Hossain2009,2023-McFadden-PRA-19-044018}. 
At the TRIUMF \gls{bnmr} facility, the \gls{slr} measurements are performed by monitoring the transient decay of the probes' spin polarization both during and following short pulses of ions (a typical length is $\Delta=$\SI{4}{\second}). 
The spin-polarization can be inferred from the $\beta$-rates detected at the two opposing (scintillator) detector pairs.
The circular polarization of the pumping laser is typically switched between pulses
resulting to alternating parallel (“+”) and anti-parallel (“-”) spin-polarizations to reduce systematic errors.~\cite{macfarlane:ZPC,macfarlane:SSNMR} 
The combined $\beta$-rates (see, e.g., ~\cite{bfit_arXiV,macfarlane:SSNMR}) result in the experimental asymmetry which is proportional to the spin-polarization:
\begin{align}
    A(t) &= \frac{1-r}{1+r} \equiv A_0 P(t),\qquad r \equiv \sqrt{\frac{(L_{+}/R_{+})}{(L_{-}/R_{-})}}
\end{align}
where $L_{+/-}$ and $R_{+/-}$ are rates in the opposing counters for the $+/-$ polarization senses, and the initial asymmetry $A_0$ depends on the properties of the geometry of the detectors (solid angle) and the intrinsic asymmetry~\cite{MacFarlane2014_InitSpinPol,Morris:ISACBook}. 

Once implanted inside the sample, the ions individually interact with the electromagnetic fields of the host lattice and their spins “relax” (depolarize) towards a dynamic equilibrium value while the beam is on, and then towards thermal equilibrium ($\approx 0$) when it is off, with a characteristic \gls{slr} relaxation time ($1/T_1$).
The magnitude of the magnetic field at the implantation site acts to slow the relaxation. Implantation depths can be varied by biasing the sample.
By recording the relaxation rates while varying the depth, the applied field, and/or the sample temperature, material properties of interest can be extracted.
In order to extract the \gls{slr} rate in \ch{Nb}, the asymmetry spectra are fit to an empirical depolarization function $p(t,t';1/T_1)$, which is convolved with the rectangular beam pulse of length $\Delta$, resulting to a bipartite asymmetry function:
\begin{align}
    A(t) &= A_0 
    \begin{cases} 
        \cfrac{R_0 \int_{0}^{t}\exp[-(t-t')/\tau] p(t,t';1/T_1) dt'}{N(t)}  &\text{for $t\leq\Delta$,}  \\
        \cfrac{R_0 \int_{0}^{\Delta}\exp[-(t-t')/\tau] p(t,t';1/T_1) dt'}{N(\Delta)}      & \text{for $t>\Delta$,}
    \end{cases} \\
    N(t) &= \int_{0}^{t}N(t,t')dt' = R_0\int_{0}^{t} \exp[-(t-t')/\tau]dt', \label{eq:A_t_fitfunc}
\end{align}
where $N(t)$ is the total number of nuclei with mean lifetime $\tau$ that are present at time $t$ after the beam has turned on and implanted at a constant rate $R_0$.

\begin{figure}[hbt!]
    \centering
    \includegraphics[width=0.5\linewidth]{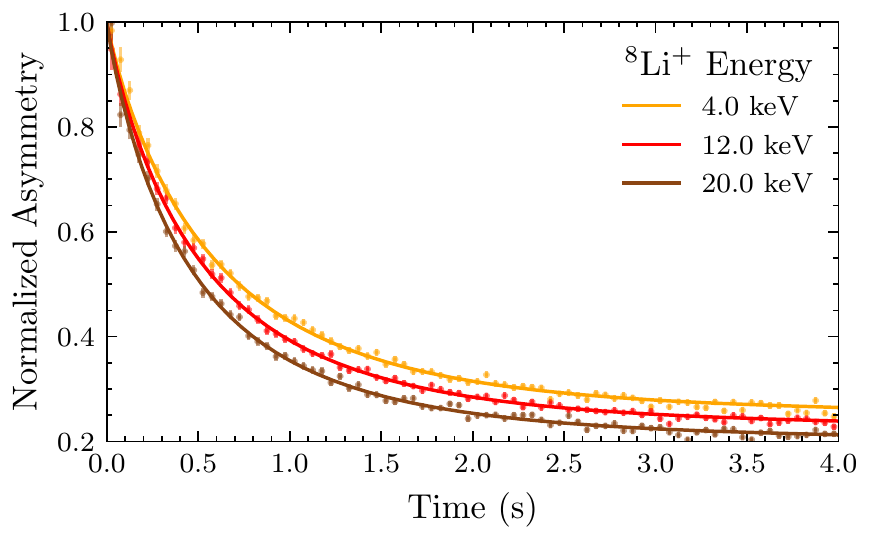}
    \caption{
        Normalized asymmetry spectra and fits to the data of the \ch{Nb} components (i.e., with background contributions subtracted) at various implantation energies for the \NbMidTAlias sample at an applied field of \SI{100}{\milli\tesla} and at $T$\SI{\sim 4.5}{\kelvin}. The increase in depolarization rate with increasing energy (i.e., subsurface stopping depth) is an indication of Meissner screening.
    } \label{fig:Asy_raw}
\end{figure}

The \LiEightPlus depolarization in thin film \ch{Nb} using a stretched exponential function has been shown to give a good description of \LiEight's spin-response at low magnetic fields~\cite{McFadden_ThinFilmNb_2023}.
The \gls{slr} becomes mono-exponential in high applied fields~\cite{Parolin:PRBNiobium}.
In our measurements, we note that a small fraction of the \ch{^{8}Li^{+}} beam stops in the surrounding cryostat components, necessitating inclusion of an additional “background” term during fitting.
This component, however, is small, slow-relaxing, and easily isolated from the signal from the \ch{Nb}.
The total depolarization model (both from the \ch{Nb} sample and the background signal) can be written as:
\begin{align}
    \label{eq:p_t_fitfunc}
    p_{z}(t, t^{\prime}) \approx f_{\ch{Nb}}\exp \left [ - \left ( \frac{t - t^{\prime}}{\TOneNb} \right )^{\beta} \right ]  + (1-f_{\ch{Nb}})\exp \left[ - \left ( \frac{t - t^{\prime}}{\TOneBG} \right ) \right],
\end{align}
where $f_{\ch{Nb}}$ is the amplitude fraction of the signal from the \ch{Nb} sample (with values of $f_{\ch{Nb}}$ ranging from 0.82 to 0.94 for \NbBase, and 0.83 to 0.93 for \NbMidTAlias), $1/\TOneNb$ is the relaxation rate of the \ch{Nb} sample, $1/\TOneBG$ is the relaxation rate of the background (with a value that is shared for all runs in each sample, $1/\TOneBG=0.1$ for \NbBase and $1/\TOneBG=0.06$ for \NbMidTAlias), and $\beta$ is the stretching exponent (shared for all runs in each sample, $\beta=0.81$ for \NbBase, $\beta=0.89$ for \NbMidTAlias).
The analysis of the asymmetry spectra is performed using the Python application bfit~\cite{bfit_paper} to yield the $\RelaxRate$ value for each data set. Examples of the depolarization during beam ON for three different ion energies are shown in \cref{fig:Asy_raw}. 
The increase in depolarization rate with increasing implantation energy is due to Meissner screening, which reduces the field as a function of depth.

As summarized in the ‘Results’ section, the analysis employs a direct mapping from the local magnetic field $B(x)$ with a corresponding Larmor frequency $\omega_i(x) = \gamma_i\ B(x)$,  to the local relaxation rate $\frac{1}{T_1}(x)$, where $\gamma_i$ is the gyromagnetic ratio of the $i$-th nuclear spin (see below).
The largest contribution to relaxation at low temperature in the superconducting state and at relatively low applied field (\SI{< 1}{\tesla}) is due to the dipolar interaction between the \LiEight probe’s nuclear spin $I=2$ and the nuclear spin $S=9/2$ of the 100\% abundant host \NbNinetyThree sample~\cite{Parolin:PRBNiobium,McFadden_ThinFilmNb_2023}. 
The dipolar relaxation rate resulting from this interaction can be expressed as a weighted sum of three Lorentzian functions, $J_n(\omega_i)$. Each of the Lorentzians is the $n$-quantum spectral density function that models the stochastic fluctuations in the transverse component of the local electromagnetic field.
This Lorentzian spectral density function is defined as:
\begin{align}
	J_{n}(\omega_{i}) &= \frac{ \tau_{c} }{1 + \omega_{i}^{2} \tau_{c}^{2} },\label{eq:spectral_density}
\end{align}
where the fit parameter $\tau_c$ is an exponential correlation time constant.

\begin{figure}[htb!]
    \centering
    \includegraphics[width=0.5\linewidth]{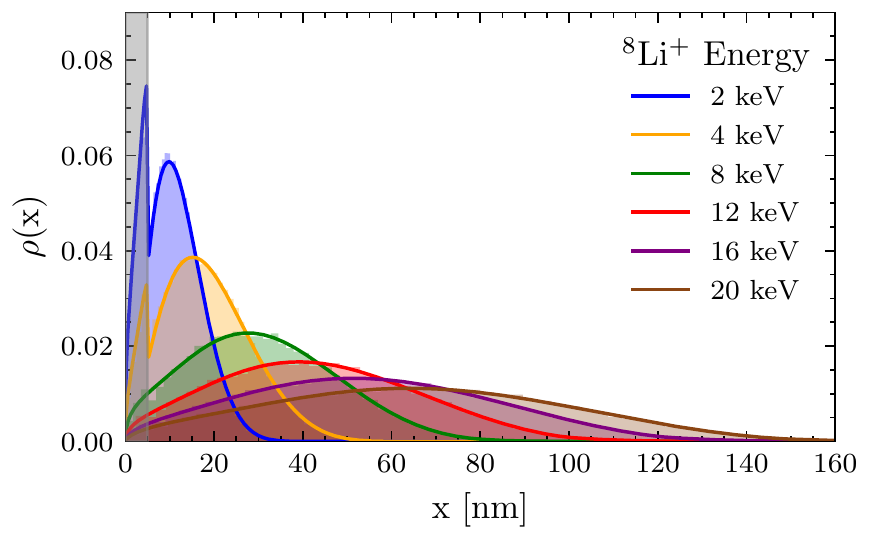}
    \caption{
        Implantation distributions $\rho(x)$ vs. stopping depth $x$ from SRIM Monte Carlo code~\cite{SRIM:Ziegler} simulation, represented as coloured histograms (with 60 bins), for various \ch{^8 Li^+} energies. 
        Solid lines are fits to a phenomenological implantation distribution function (for details, see\,\cite{McFadden_ThinFilmNb_2023}).
        The grey area indicates the commonly present \ch{Nb2O5} native-oxide layer with typical thickness of \SI{\sim 5}{\nano\meter}.}
    \label{fig:SRIM_profile}
\end{figure}

The local relaxation rate due to the dipolar interaction can be expressed as~\cite{Mehring:PHRNMRS-1983, McFadden_ThinFilmNb_2023}:
\begin{align}
    \frac{1}{T_1} (x) &= \frac{1}{T_1} \left [ B(x) \right ]\nonumber\\
                    &=  \gLi \gNb B_{d}^{2} \left \{ \frac{1}{3} J_{0}( \omega_{S} - \omega_{I} ) + J_{1}( \omega_{I} ) + 2 J_{2}( \omega_{S} + \omega_{I} ) \right \} ,\label{eq:SLR_rate_heteronuclear}
\end{align}
where $B_d$ is the magnitude of the fluctuating dipolar field at the \LiEight site) and is another fit parameter, $\gLi$ is the gyromagnetic ratio of the \LiEight probe nuclei, and $\gNb$ ($=2\pi\times$ 10.439565(3388) \unit{\mega\hertz/\tesla})~\cite{2019-Stone-INDC-NDS-0794} is the gyromagnetic ratio of the \NbNinetyThree host nuclei.
To reconstruct the local field, modelled as \cref{eq:B_vs_x} or \cref{eq:B_vs_x-vortex}, we perform a chi-square minimization between the measured SLR rate and an average of the local SLR rate from \cref{eq:SLR_rate_heteronuclear} using the implantation distribution for a given energy:
\begin{align}
    \left\langle \frac{1}{T_1}\right\rangle_E = \int_0^{\infty} \rho_E(x) \frac{1}{T_1}(x) dx.\label{eq:Ave_Relax}
\end{align}
Here $\rho_E(x)$ is the implantation distribution for ions with implantation energy $E$.
These distributions can be accurately simulated using Monte Carlo codes such as SRIM~\cite{SRIM:Ziegler} and are shown in \cref{fig:SRIM_profile}.
The fit to the distribution using a phenomenological distribution function (described in detail in \, \cite{McFadden_ThinFilmNb_2023}) is also shown.
A \SI{5}{\nano\meter} layer of \ch{Nb2O5} on a \ch{Nb} substrate is assumed in the simulation to represent a typical \ch{Nb} system (see, e.g., \cite{Halbritter1987_OxideThick, kolb2023midt, Wenskat2022_surfmag}).

\bibliography{references}

\section*{Acknowledgements}

The authors of this work would like to thank:
M.~Cervantes, D.~O.~Rosales for assistance with sample preparation, as well as TRIUMF CMMS, SRF, DAQ, Operations and Beam Delivery groups for excellent technical support before and during the experiment.
Funding was provided through a Research Tools and Infrastructure grant [SAPEQ-2015-00005] and Discovery grant [SAPIN-2019-00040] and [SAPPJ-2020-00030] from the Natural Sciences and Engineering Research Council of Canada (NSERC).  

\section*{Author contributions statement}

\Edward  and \Bob wrote the main manuscript with minor contributions from \Ryan, and all authors reviewed the manuscript.
\Edward and \Philipp prepared the samples.
\Edward, \Bob, \Asad, and \Ryan designed the research.
\Edward, \Bob, \John, \Asad, \Sarah, \Gerald, \Victoria, \Derek, \Andrew, \Ruohong, and \Suresh  contributed to the \bNMR measurements.
\Edward and \Bob performed the analysis.  
Fitting routines were developed by \Edward and \Bob based on original routines developed by \Ryan
\Bob, \Tobi, \Kiefl, and \Andrew supervise the contributing participants of this experiment.
\Bob, \Kiefl, and \Tobi contribute to the funding acquisition.

\section*{Data availability statement}

The data and fitting routines supporting the findings of this study are available from the corresponding authors upon reasonable request. 
Raw data from the \gls{bnmr} experiments (experiment number M1963) are available for download from: \url{https://cmms.triumf.ca}.

\section*{Additional information}

\textbf{Competing interests:} The authors declare no competing financial interests.

\end{document}


\flushbottom
\maketitle

\section*{\SIBint}

In order to simplify our analysis in the mixed phase (Case B), we limit the number of added fit parameters by relating $\BInt$ to $\BSurf$ via the normalized enhancement factor $\mathcal{N}$.
First, we assume a pin-free superconductor with a constant equilibrium vortex distribution (and vortex density, $n_v(x) = \tilde{n}$), resulting to a constant internal field $\BInt = \mu_0(\HInt + M)$.
Furthermore, we also assume a constant effective magnetization $M$ within the volume probed by the \LiEightPlus beam.
Using these two assumptions, an implicit equation of the internal field $\HInt$ can be obtained via the constitutive relation $M(\BApp) = -\abs{\chi(\BApp)}\HInt$ (i.e. the diamagnetic properties of the superconductor), and by defining a geometry dependent effective demagnetization factor ($D$):
\begin{align}
    \HInt &= \HApp - D M(\HApp) \nonumber \\
           &= \HApp + D \abs{\chi(\BApp)} \HInt \nonumber \\
    \HInt &= \frac{\HApp}{1 - D\abs{\chi(\BApp)}}.\label{eq:HInt-chi}
\end{align}
Applying the boundary condition for the continuity of parallel $H$-field, i.e., $\HInt=\HSurf$, results to:
\begin{align}
    \BSurf &= \mu_0\HSurf, \nonumber \\
           &= \frac{\BApp}{1 - D\abs{\chi(\BApp)}}. \label{eq:BSurf-chi}
\end{align}
The equation above is the general expression describing the enhancement of $\BSurf$ which depends on the diamagnetic property of a superconductor (see, e.g., \cite{Poole2014_Ch5}).

Equating our definition of $\BSurf$ which utilizes the normalized enhancement factor $\mathcal{N}(\BApp)$ and \cref{eq:BSurf-chi}, we can obtain a mapping between ($\mathcal{N}$,$\mathcal{E}_M$) as described in the text with ($\abs{\chi}$,$D$):
\begin{align}
    \BSurf/\BApp &= \mathcal{E},\nonumber \\
    \frac{1}{1-\abs{\chi}D} &= 1 + \mathcal{N}(\mathcal{E}_M - 1). \label{eq:BSurf-equivalence}
\end{align}
The right side of \cref{eq:BSurf-equivalence} is the definition of the enhancement factor ($\mathcal{E}$) described in the text, while the left side is the general equation for $\BSurf$ for a superconductor with spatially constant magnetization and a demagnetization factor $D$ (see e.g., \cite{Poole2014_Ch5}).
The effective demagnetization factor $D$ can be related to $\mathcal{E}_M$, the $\mathcal{E}$ value in the Meissner phase, by requiring $\abs{\chi} = \mathcal{N} = 1$ to obtain:
\begin{align}
    D = \frac{\mathcal{E}_M - 1}{\mathcal{E}_M}.\label{eq:demag-enhanceMeissner}
\end{align}
The absolute susceptibility can also be expressed in terms of the enhancement factor $\mathcal{E}$, $\mathcal{E}_M$ in the Meissner phase, and the normalized enhancement factor $\mathcal{N}$ as:
\begin{align}
    \abs{\chi} &= \left(\frac{\mathcal{E} - 1}{\mathcal{E}}\right) \frac{1}{D},\nonumber\\
              &= \frac{\mathcal{E}_M}{\mathcal{E}}\mathcal{N}. \label{chi-N-relation}
\end{align}
The assumption of spatially constant $M$ and $D$ do not apply for non-ellipsoidal superconductor in a strict physical sense (see, e.g., \cite{Brandt:GeometPaper}). 
They are defined here, however, as the effective (constant) quantities to give $\mathcal{E}_M$ equal to the known calculated value from finite element simulation.

Using the aforementioned definitions, we can now relate the $\BInt$ to $\BApp$ as:
\begin{align}
    \BInt &= \mu_0 (\HInt + M), \nonumber \\
        &= \mu_0\HInt \left(1 - \frac{\mathcal{E}_M}{\mathcal{E}}\mathcal{N} \right),\nonumber\\
        &= \mu_0\HApp \mathcal{E} \left( \frac{\mathcal{E}-\mathcal{E}_M\mathcal{N}}{\mathcal{E}} \right), \nonumber\\
        &= (1-\mathcal{N})\BApp, \label{eq:BInt-BApp}
\end{align}
which is the simple linear relationship between $\BInt$ and $\BApp$ quoted in the text.
Therefore, the constant $\BInt\left[\mathcal{N}(\BApp) \right]$ is calculated rather than varied as a fit parameter, whereas $\mathcal{N}(\BApp)$ is a fit parameter extracted at each applied field.

\bibliography{references}